\documentclass[useAMS,usenatbib]{mn2e}
\usepackage{graphicx,color,epsfig}

\newcommand{\be}{\begin{equation}}
\newcommand{\ee}{\end{equation}}

\newcommand{\apj}{ApJ}

\newcommand{\mnras}{MNRAS}
\newcommand{\aap}{A\&A}
\newcommand{\araa}{ARA\&A}
\newcommand{\apjl}{ApJL}

\def\ltsima{$\; \buildrel < \over \sim \;$}
\def\simlt{\lower.5ex\hbox{\ltsima}}
\def\gtsima{$\; \buildrel > \over \sim \;$}
\def\simgt{\lower.5ex\hbox{\gtsima}}

\newcommand\rfc{R_{\rm fc}}
\newcommand\mfc{M_{\rm fc}}
\newcommand\lfc{L_{\rm fc}}

\def\msun{{\,{\rm M}_\odot}}

\def\del#1{{}}

\title[Grain sedimentation in giant embryos]{Grain sedimentation inside giant
  planet embryos.}

\author[S. Nayakshin]{Sergei Nayakshin \\ Department of Physics \& Astronomy,
  University of Leicester, Leicester, LE1 7RH, UK}

\begin{document}

\date{Accepted 2008 ?? ??. Received 2008 ?? ??; in original form 2008 05 ??}

\pagerange{\pageref{firstpage}--\pageref{lastpage}} \pubyear{2008}

\maketitle

\label{firstpage}

\begin{abstract}
In the context of massive fragmenting protoplanetary discs, \cite{Boss98}
suggested that grains can grow and sediment inside giant planet embryos formed
at $R\sim 5$ AU away from the star. Several authors since then criticised the
suggestion. Convection may prevent grain sedimentation, and the embryos cannot
even form so close to the parent star as cooling is too inefficient at these
distances. Here we reconsider the grain sedimentation process suggested by
\cite{Boss98} but inside an embryo formed, as expected in the light of the
cooling constraints, at $R\sim 100$ AU. Such embryos are much less dense and
are also cooler. We make analytical estimates of the process and also perform
simple spherically symmetric radiation hydrodynamics simulations to test these
ideas. We find that convection in our models does not become important before
a somewhat massive ($\sim$ an Earth mass, this is clarified in a followup
paper) solid core is built. Turbulent mixing slows down dust sedimentation but
is overwhelmed by grain sedimentation when the latter grow to a centimetres
size.  The minimum time required for dust sedimentation to occur is a few
thousand years, and is a strong function of the embryo's mass, dust content
and opacity. An approximate analytical criterion is given to delineate
conditions in which a giant embryo contracts and heats up faster than dust can
sediment. As \cite{BossEtal02}, we argue that core formation through grain
sedimentation inside the giant planet embryos may yield an unexplored route to
form giant gas and giant ice planets. The present model also stands at the
basis of paper III, where we study the possibility of forming terrestrial
planet cores by tidal disruption and photoevaporation of the planetary
envelope.
\end{abstract}


\section{Introduction}\label{intro}

The Gravitational Instability (GI) model for giant planet formation
\citep[e.g.,][]{Bodenheimer74,Boss97} is one of the two well known models for
planet formation \citep[e.g.,][]{Wetherill90}. The model has been thought to
be unable to produce solid cores observed in giant planets in the Solar System
\citep[e.g.,][]{Fortney09} until \cite{Boss98,BossEtal02} demonstrated that
dust growth and sedimentation may realistically occur inside gaseous
protoplanets. However, \cite{WuchterlEtal00} argued that giant planet embryos
may become convective within the first $\sim 100$ years of their evolution,
and that convection may inhibit grain sedimentation. \cite{HelledEtal08,HS08}
have recently confirmed that convection may indeed be a serious obstacle to
solid core formation inside the giant planet embryos. These authors obtained
much smaller cores than the model of \cite{Boss98} predicts.  More recently,
the GI model for giant planet formation has been criticised on the grounds
that the cooling time at the inner, e.g., $R\simlt 10$ AU disc, is not short
enough to permit a continuous gravitational contraction of gaseous clumps
\citep[e.g.,][]{Rafikov05,Rice05}.

Whilst this criticism of \cite{Boss98} ideas appears to be relevant, there are
now extrasolar giant planets observed at distances out to hundreds of AU from
their parent star, where the core accretion model
\citep[e.g.,][]{Weiden80,Wetherill90} for planet formation may not have been
able to create solid cores before the gaseous disc dissipates. It seems rather
likely that the GI model is needed to explain these planets. Simulations of
large (e.g., tens of AU) discs that employ realistic cooling in the optically
thick regime do show formation of gaseous clumps with masses of a few to a few
tens of Jupiter masses \citep[e.g.,][]{SW08,Meru10}, which presumably cool and
contract into giant planets or brown dwarfs.

It is timely to reconsider the gravitational disc instability model taking
into account the updated constraints on the disc fragmentation
\citep[e.g.,][]{Rafikov05}. To investigate this complicated and non-linear
problem properly, we divide presentation of our work into several papers --
(1, this paper) before and (2, Nayakshin 2010b, to be submitted) after the
solid core formation, and (3, Nayakshin 2010c, submitted to MNRAS letters)
interplay of the processes occurring inside the embryo with external influences
of the parent disc and the star.

The goal of this paper (referred in later papers as ``paper I'') is to study
dust growth and sedimentation inside the giant embryo but excluding the solid
core formation. We limit our study to isolated non-rotating giant planet
embryos or first cores, the terms that we use inter-exchangeably below. While
the study of such a limited problem may appear too academical, we shall use
these results extensively in papers II and III. There is also a possibility
that dust sedimentation inside first cores may also be relevant to the process
of low mass star formation. Therefore we hope that analytical results
presented here may be useful for future researchers in more than one field.

The gist of the ``new'' disc fragmentation constraints is that the giant
planet embryo must start off much less dense than previously assumed. Namely,
\cite{Boss98} assumed that the protoplanet has a mean density of order $\sim
10^{-8}$ g cm$^{-3}$. Such high a density is needed to overcome the tidal
fields from the parent star at several AU distances. However, at $\sim 100$
AU, the initial embryo density of order $\sim 10^{-13}$ g cm$^{-3}$ is
expected (see paper II for detail). Optical depths of such embryos are much
less than of those studied by
\cite{Boss98,WuchterlEtal00,HelledEtal08}. Therefore we find that the giant
planet embryo remains radiative long enough to permit dust growth and
sedimentation. In fact, convection becomes important only after the solid core
assembly as the energy release by the solid core rises to as much as $10^{28}
- 10^{29}$erg s$^{-1}$ (paper II).

The paper is organised as following. In \S \ref{sec:formation}, we introduce
the so-called ``first cores'' and argue that the initial state of giant planet
embryos should be similar to that of the first cores. In \S
\ref{sec:int_evolution} we study the internal evolution of {\em isolated}
first cores, defining the maximum gas accretion rate onto the first core under
which the core can be considered isolated. We then attempt to carry out the
calculations analytically as far as we can, building a simple analytical model
to follow the cooling and contraction of the first cores, and grain growth and
sedimentation inside them. We also consider the influence of turbulent mixing
and grain shattering via high speed collisions that may delay the onset of
grain sedimentation.

In later parts of the paper, \S \ref{sec:num_isolated} and \S
\ref{sec:num_results}, we use a simple spherically symmetric radiation
hydrodynamics code with a two-fluid approach to capture grain growth and
dynamics to allow for a simultaneous evolution of the gas and the dust. We
find higher opacity, larger metalicity and lower mass first cores to be the
most promising sites of grain sedimentation. Turbulent mixing mainly slows
down rather than forbids grain sedimentation; although in corners of the
parameter space it may be crucially important. We conclude that whether the
grains sediment inside the first core or not mainly depends on whether it
contracts and hence heats to the grain vaporisation temperature faster than
the grains can grow. Finally, in the Discussion section we overview the main
results of our work, compare those to previous work, and consider implications
for the field of planet formation. We argue that the planet formation
community should take a detailed look at the giant instability disc model
supplemented by a realistic internal evolution for the giant embryo, and also
embryo dynamics in the parent disc.

\section{First cores and giant planet embryos}\label{sec:formation}

\subsection{First cores in studies of star formation}

Stars form from larger gas reservoirs due to gravitational contraction of the
latter. The simplest case of a spherical, uniform, non-rotating, and initially
at rest, constant density cloud is well known \citep{Larson69}.  The
collapsing gas is initially isothermal, as cooling is very effective. However,
as the innermost region of the collapsing gas fragment becomes denser, the
free fall time there becomes shorter than the cooling time of the gas
\citep{MI99}. The region switches from an isothermal to an adiabatic
behaviour. Heated up, the gas is able to stop the further collapse by thermal
pressure forces. The central part of the region settles into a hydrostatic
balance. This region is logically named ``the first core''.

The minimum mass of the first cores is given by the ``opacity limit''
\citep{Rees76,Low76} of a few Jupiter masses \citep[although][note that the
  term is somewhat physically misleading]{MI99}, and the size is $\sim 10^{14}
$ cm. The conditions in a typical {\em non rotating} Solar mass molecular gas
cloud puts the first cores in a situation where the mass supply rate is
continuous and relatively high ($\sim$ a few $\times 10^{-5} \msun$
yr$^{-1}$). In this case new matter lands on the first core so fast that it
does not have time to cool, and therefore the behaviour of the gas nearly
adiabatic \citep{Larson69,Masunaga98,Masunaga00}. They contract and heat up
only due to more and more mass piling up on them from the parent gas
cloud. When their mass exceeds about 50 Jupiter masses, the cores reach the
temperature of about $2000 K$ and the gas density about $10^{-8}$ g
cm$^{-3}$. Dissociation of molecular hydrogen then ensues. This provides an
additional energy sink and initiates a ``second collapse'' which finally turns
the first core into a true proto-star as compact as a few Solar radii.

To understand why the first cores do not collapse further until they reach a
mass of about $M_{\rm max} = 0.05 \msun$ 
\citep{Masunaga98,Masunaga00}, consider the virial temperature of the first
core. As will be shown latter (equation \ref{tvir_core}), an adiabatically
evolving first core has the virial temperature of about
\begin{equation}
T_{\rm vir} = \frac{GM_{\rm fc}\mu}{3 k_B R_{\rm fc}} \sim 150 \;\hbox{K}\;
\left(\frac{\mfc}{0.01 \msun}\right)^{4/3}\;,
\label{tvir_est}
\end{equation}
where $\mfc$ and $R_{\rm fc}$ are the mass and the radius of the first core,
respectively.  The core becomes hot enough for hydrogen to disassociate only
when it reaches the mass of about 50 Jupiter masses.

\subsection{Giant planet embryos are similar to first cores}\label{sec:why}

Consider now the properties of giant planet embryos at the moment of their
formation in a protoplanetary disc.  In order to fragment gravitationally, a
massive gas disc must satisfy two criteria. First of all, it must be
massive/dense enough. In particular, the \cite{Toomre64} Q-parameter must be
less than unity:
\begin{equation}
Q = \frac{c_s \Omega}{\pi G \Sigma} \leq 1\;,
\label{toomre}
\end{equation}
where $c_s$ is the gas isothermal sound speed; $\Omega = (G M_*/R_p^3)^{1/2}$ is
Kepler's angular frequency; $M_*$ is the mass of the star, assumed to be
larger than that of the disc, and $\Sigma = 2 H \rho$ is the surface density
of the disc. Here $\rho$ is the disc vertically averaged density, and $H = c_s
\Omega^{-1}$ is the disc vertical scale-height. Rearranging equation
\ref{toomre}, we see that at $Q=1$ the disc mean density is equal to
the tidal density of the star,
\begin{equation}
\rho_{\rm t} = \frac{M_*}{2 \pi R_p^3} \; \approx 10^{-13} \;\hbox{g
  cm}^{-3}\; \frac{M}{\msun}R_2^{-2}\;,
\label{rhotd}
\end{equation}
to within a factor of order unity, where $R_2 = R_p/100$ AU.

The second condition for disc fragmentation is the condition on the rate of
cooling. Expressed in terms of the cooling time $t_{\rm cool}$, the
requirement states \citep{Gammie01,Rice05}:
\begin{equation}
t_{\rm cool} \simlt 3 \Omega^{-1}\;.
\label{tclcr}
\end{equation}
This implies that the critical radiative cooling rate per unit gram of disc
material  is
\begin{equation}
\Lambda_{\rm crit} = \frac{e}{t_{\rm cool}} \approx \frac{c_s^2 \Omega}{3(\gamma-1)}\;,
\label{ccr}
\end{equation}
where $\gamma$ is the specific heats ratio for gas. At the same time, the
compressional heating rate for gravitational contraction is \citep{MI99}
\begin{equation}
\Lambda_{\rm grav} = c_s^2 \left(4\pi G \rho\right)^{1/2} = 2^{1/2}\; c_s^2 \Omega\;,
\label{ccr}
\end{equation}
where we have used equation \ref{rhotd} to eliminate $\rho$. It is now clear
that $\Lambda_{\rm crit} \approx \Lambda_{\rm grav}$ at the point of disc
fragmentation\footnote{This statement is accurate within a factor of order
  unity, e.g., the same order of accuracy to which the vertically integrated
  accretion disc equations \citep{Shakura73} are to be trusted.}. This is
intuitively correct, as otherwise the disc could heat up faster than it could
fragment, increasing $c_s$ and returning itself back to stability (e.g., $Q >
1$).

We have shown above that a fragmenting self-gravitating gas disc satisfies
$\Lambda_{\rm crit} \approx \Lambda_{\rm grav}$. This is also the condition
that sets the ``opacity limit for fragmentation'' in star formation
\citep{Rees76,Low76,Masunaga98,MI99}, which yields a minimum mass for
fragmentation of a few Jupiter masses. This is of course not a coincidence as
in both cases gravitational collapse is stopped if the gas is unable to
continue cooling rapidly enough.

Therefore the properties of the gaseous clumps in the disc at their inception
should be similar to that of the first cores of a few Jupiter masses. As with
the first cores, the clumps may gain more mass from the disc. However, the
giant planet embryos in the disc have to compete for gas with other embryos,
as the disc is likely to fragment on many rather than one clump. Collisions
between clumps may unbind the clumps partially or completely. Furthermore, it
is well known that above the ``transition'' mass $M_t \approx 2 M_* (H/R)^3$
\citep{BateEtal03} the planet opens up a gap in the disc due to gravitational
torques between the planet and the disc, and that the gap opening curtails
gas accretion onto the planet by orders of magnitude
\citep[][]{LubowEtal99,BateEtal03}. Giant planet embryos with mass of the
order of 10 Jupiter masses are in this gap-opening regime as long as $H/R
\simlt 0.2$ for a Solar mass star.

Therefore, for simplicity, we study the evolution of planet embryos at a
constant mass in this paper.  The mass of the gaseous clump is a free parameter
below. We also use the terms ``first core'' and ``giant planet embryo''
inter-exchangeably throughout.

\section{Internal evolution of first cores}\label{sec:int_evolution}

\subsection{Physical properties of adiabatic first cores}\label{sec:1st_analytical} 

\cite{Masunaga98} show that the structure of the first cores can be well
approximated by a polytropic sphere, with the polytropic constant fixed by the
gas temperature and the density $\rho_{\rm ad}$ at which the gas switches from
the isothermal to the adiabatic behaviour. The typical values of these are
often taken in the literature as $T_{\rm init} = 10$ K and $\rho_{\rm ad} =
10^{-13}$ g cm$^{-3}$.  However, $\rho_{\rm ad}$ does depend on the opacity of
the material and the ambient gas temperature $T_{\rm init}$
\citep{MI99}. Following these authors, we take opacity to be dominated by
dust, in the functional form
\begin{equation}
\kappa(T) = \kappa_0 \left(\frac{T}{10}\right)^\alpha\;,
\label{kappa0}
\end{equation}
where $\kappa_0 $ is the opacity at $T= 10$ K, and reasonable values of
$\alpha$ are thought to be between 1 and 2. With this, the adiabatic density
threshold is best described by the following \citep{MI99}:
\begin{equation}
\rho_{\rm ad} = 5\times 10^{-13} \kappa_{*}^{-2/3} T_1^{\frac{4-2\alpha}{3}}
\;\hbox{g cm}^{-3}\;,
\label{mi99}
\end{equation}
where $\kappa_{*} = \kappa_0/0.01$ and $T_1 = T_{\rm init}/10$K.

\cite{Masunaga98} give expression for $\rfc$ as a function of the central
density. Eliminating the density in favour of the sphere's mass and radius, we
obtain
\begin{equation}
\rfc = 17.5 \; \hbox{AU}\; m_1^{-1/3} T_1 \rho_{-13}^{-2/3}\;,
\label{rfc}
\end{equation}
where $\rho_{-13} = 10^{13} \rho_{\rm ad}$, $T_1 = T_{\rm init}/10$ and $m_1=
\mfc/0.01 \msun$. Eliminating the critical density by using equation
\ref{mi99}, we get 
\begin{equation}
\rfc = 6.0 \;\hbox{AU}\; m_1^{-1/3} T_1^{\frac{1+4\alpha}{9}}\kappa_{*}^{4/9}\;.
\label{rfc2}
\end{equation}
Note that the temperature dependence in this expression is at most linear, and
the opacity enters in an even weaker power. The initial size of the first core
is hence varies rather little. The mean density of the first core is defined
by
\begin{equation}
\rho_{\rm mean} = \frac{3 \mfc}{4\pi \rfc^3} = 6.6\times 10^{-12}\; \hbox{g
  cm}^{-3}\; m_1^{2} T_1^{-\frac{1+4\alpha}{3}}\kappa_{*}^{-4/3}\;,
\label{rho_mean}
\end{equation}
and is a very strong function of the mass of the first core.  The temperature
of the first core is of the order of the virial temperature, which we estimate
to be
\begin{equation}
T_{\rm vir} = \frac{1}{3} \frac{G \mfc \mu}{k_B \rfc} = 146 \; \hbox{K}\;
m_1^{4/3} T_1^{-\frac{1+4\alpha}{9}}\kappa_{*}^{-4/9}\;.
\label{tvir_core}
\end{equation}
From this it follows that the sound speed in the first core is, $c_s =
\sqrt{kT/\mu} \approx \sqrt{G\mfc/\rfc}$,
\begin{equation}
c_s = 0.7 \;\hbox{km s}^{-1}\; m_1^{2/3} T_1^{-\frac{1+4\alpha}{18}}\kappa_{*}^{-2/9}
\end{equation}

The column depth of the first core is
\begin{equation}
\Sigma_{\rm fc} = \frac{\mfc}{\pi \rfc^2} = 1.18\times 10^3 \; \hbox{g cm}^{-2} \;
m_1^{5/3} T_1^{\frac{2+8\alpha}{9}}\kappa_{*}^{-8/9}\;,
\label{sigma_core}
\end{equation}
and the optical depth of the first core is
\begin{equation}
\tau_{\rm core} = \kappa(T_{\rm vir}) \Sigma_{\rm fc} = 11.8 \times
(14.6)^\alpha m_1^{\frac{5+4\alpha}{3}} \phi_1(T_1,k_{*})\;,
\end{equation}
where $\phi_1(T_1,\kappa_{*}) = T_1^{\frac{2+7\alpha -4 \alpha^2}{9}}
\kappa_{*}^{\frac{1-4\alpha}{9}}$. From this expression it is obvious that
the first cores are always optically thick. The radiative luminosity of the
first core is
\begin{equation}
\lfc \approx \frac{4 \pi \rfc^2 \sigma_B T_{\rm vir}^4}{\kappa \Sigma_{\rm
    fc}}\;,
\label{lrad}
\end{equation}
or numerically, 
\begin{equation}
\lfc = 2.2\times 10^{32} (14.6)^{-\alpha} m_1^{\frac{9-4\alpha}{3}}
\phi_2(T_1,\kappa_{*}) \;,
\end{equation}
where function $\phi_2(T_1,\kappa_{*}) = T_1^{(4-15\alpha+4\alpha^2)/9} \;
\kappa_{*}^{-1+4\alpha/9}$ is a cumbersome looking power-law function which
does not vary much for $\alpha$ between 1 and 2, however. For example, for
$\alpha = 1$, $\phi_2(T_1,\kappa_{*}) = T_1^{7/9} \kappa_{*}^{-5/9}$,
whereas for $\alpha = 2$, $\phi_2(T_1,\kappa_{*}) = T_1^{10/9}
\kappa_{*}^{-1/9}$. Nevertheless, it does become quite inconvenient to write
down expressions for a general value of $\alpha$, and so we shall present the
results only for either $\alpha=1$ or $\alpha=2$ below.  For convenience of
future reference,
\begin{equation}
\lfc = 
\cases{1.5 \times 10^{31} m_1^{5/3} T_1^{7/9} k_{*}^{-5/9}\qquad \hbox{for}\; \alpha = 1 \cr
1.0 \times 10^{30} m_1^{1/3} T_1^{10/9} k_{*}^{-1/9} \qquad \hbox{for}\; \alpha = 2\;.\cr}
\label{lfc0}
\end{equation}
Note that the luminosity is lower for $\alpha=2$ than for the other case. This
is due to a higher opacity of the first core, e.g., the larger exponent in the
opacity law.  The binding energy of the polytropic sphere,
\begin{equation}
E_{\rm bind} = \frac{3}{5} \frac{G \mfc^2}{\rfc} = 1.8 \times 10^{41} \; \hbox{erg}\;
m_1^{7/3} T_1^{-\frac{1+4\alpha}{9}} \kappa_{*}^{-4/9}\;,
\label{ebind_fc}
\end{equation}
 allows us to calculate the cooling time of the sphere as
\begin{equation}
t_{\rm cool} = \frac{E_{\rm bind}}{\lfc}\;.
\label{tcool}
\end{equation}
For the smaller value, $\alpha=1$, 
\begin{equation}
t_{\rm cool} = 380 \; \hbox{years}\; m_1^{2/3} T_1^{-4/3} \kappa_{*}^{1/9}\;,
\label{tcool1}
\end{equation}
whereas for $\alpha=2$, 
\begin{equation}
t_{\rm cool} = 5700 \; \hbox{years}\; m_1^{2} T_1^{-19/9} \kappa_{*}^{-1/3}\;.
\label{tcool2}
\end{equation}

Another useful time scale is the dynamical time of the core, which is of the
order of the free fall time for the core:
\begin{equation}
t_{\rm dyn} = \left(G \rho_{\rm mean}\right)^{-1/2} \approx 50 \;\hbox{yrs} \;
m_1^{-1}T_1^{\frac{1+4\alpha}{6}}\kappa_{*}^{2/3}
\label{tdyn}
\end{equation}
Note that the cooling time is indeed longer than dynamical time, as should be
for a hydrostatic gas configuration.

\subsection{Gas-starved regime for the first cores}\label{sec:starved}

Having reviewed the adiabatic evolution of the first cores, we can now estimate
when cooling of the cores becomes important. As the core gains (or perhaps
looses mass), the adiabatic temperature time derivative is
\begin{equation}
\left(\frac{d T_{\rm vir}}{dt}\right)_{\rm ad} = \frac{\partial T_{\rm
      vir}}{\partial \mfc} \frac{d \mfc}{dt} = \frac{4}{3}\;T_{\rm vir}
  \frac{d \ln \mfc}{dt}\;,
\label{dTdt_ad}
\end{equation}
where we used equation \ref{tvir_core}.
On the other hand, the temperature derivative due
to radiative cooling is
\begin{equation}
\left(\frac{d T_{\rm vir}}{dt}\right)_{\rm rad} = \frac{T_{\rm vir}}{t_{\rm cool}}
\label{dTdt_radc}
\end{equation}
(see also equation \ref{dTdt} below). We define a critical ``radiative'' first
core accretion or decretion rate as the one for which $|(dT_{\rm vir}/dt)_{\rm
  rad}| = (dT_{\rm vir}/dt)_{\rm ad}$. For mass gain or loss rate below this
critical rate, the radiative cooling is the dominant driver of the cloud's
thermal evolution. From equation \ref{dTdt_radc} and \ref{dTdt_ad} we find
\begin{equation}
\left|\frac{d \mfc}{dt}\right|_{\rm rad} = \frac{3 \mfc}{4 t_{\rm cool}}= 
\label{dMdt_crr}
\end{equation}
\begin{equation}
= 
\cases{2 \times 10^{-5} \msun \hbox{yr}^{-1}\; m_1^{1/3} T_1^{4/3} k_{*}^{-1/9}
  \qquad \hbox{for}\; \alpha = 1 \cr
1.3 \times 10^{-6} \msun \hbox{yr}^{-1}\; m_1^{-1} T_1^{19/9} k_{*}^{1/3} \qquad \hbox{for}\; \alpha = 2\cr}
\label{dMdt_crr_num}
\end{equation}
Note further that as the first cores contract, their cooling time $t_{\rm
  cool}$ increases (\S \ref{sec:evolution} below), and hence the critical
radiative accretion rate actually drops with time from the value found in
equation \ref{dMdt_crr_num}.

We can now compare the critical radiative accretion rates with the accretion
rates of the first cores in the context of a Solar mass gas cloud collapsing,
for which the first cores accumulate mass at the rate $d\mfc/dt \sim 10^{-5}
\msun$ yr$^{-1}$ \citep{Larson69,Masunaga98,Masunaga00}. These rates are
higher than the critical radiative accretion rates, and therefore the first
cores should indeed evolve quasi-adiabatically in these conditions.

In the opposite case, $|d\mfc/dt| \ll (d\mfc/dt)_{\rm rad}$, the evolution of
the first cores will be governed mainly by their radiative cooling. This is
the case that we henceforth study in this paper.

\subsection{Cooling of isolated cores}\label{sec:evolution}

Staying within the analytical approach, we calculate the cooling evolution of
the first cores, assuming that they remain polytropic spheres. The solutions
obtained in this way will be later compared to a numerical calculation that
does not use this assumption.

We obtain the solution by solving the equation $d E_{\rm bind}/dt = \lfc$,
where the plus sign is used because we defined the binding energy to be
positive. For convenience, we introduce dimensionless temperature $\tilde
T(t)$ of the first core as the ratio of the current virial temperature $T(t)$
to the initial virial temperature of the core at formation, as given by
equation \ref{tvir_core}. We also introduce a constant for a given $\mfc$,
$t_0 = t_{\rm cool}(0)$, which is the cooling time of the first core at
formation, given by equation \ref{tcool}.

Equation \ref{lrad} shows that luminosity of the first core changes with time
as 
\begin{equation}
L_{\rm fc} \propto \rfc^4 T^{4-\alpha} \propto \tilde T^{-\alpha}\;;
\label{lscale}
\end{equation}
where we used the fact that $\rfc T\propto$ is a constant for a given $\mfc$.
Since the binding energy scales as $E_{\rm bind} \propto T(t)$ at a fixed
$\mfc$, the equation for the evolution of the dimensionless temperature of the
first core can be written as
\begin{equation}
\frac{d \tilde T}{d t} = \frac{\tilde T^{-\alpha}}{t_0}\;,
\label{dTdt}
\end{equation}
which is solved as
\begin{equation}
\tilde T(t) = \left[1 + \left(1+\alpha\right)\frac{t}{t_0}\right]^{\frac{1}{1+\alpha}}\;.
\label{Toft}
\end{equation}
Using this, we find for the cooling time evolution
\begin{equation}
t_{\rm cool}(t) = t_{\rm cool}(0) + (1+\alpha)t\;,
\label{tcev}
\end{equation}
which in the limit $t\gg t_{\rm cool}(0)$ shows that the cooling time is
always $1+\alpha$ times the current time. Note that this implies that the
first core's cooling becomes more inefficient as it contracts. The luminosity
of the core evolves according to
\begin{equation}
L_{\rm fc}(t) = L_{\rm fc}(0) \left[1 +
  \left(1+\alpha\right)\frac{t}{t_0}\right]^{\frac{\alpha}{1+\alpha}}\;,
\label{lum_of_t}
\end{equation}
where $L_{\rm fc}(0)$ is the initial luminosity of the core, given by equation
\ref{lfc0}. Note that equation \ref{lum_of_t} is an increasing function of
time, so that the first cores do become brighter as they contract, but the
binding energy of the core increases with $t$ even faster, and this is why the
cooling time is a growing function of time as well.

We can now define the time scale $t_{\rm vap}$ defined as the time needed for
the first core to heat up to the temperature $T_{\rm vap}$ at which grains
vaporise (see \S \ref{sec:num_ggrowth}; $T_{\rm vap} \sim 1200 - 1400 K$,
depending on the size of the grains).  This time scale is obtained by solving
\begin{equation}
T_{\rm vap} = T_{\rm vir} \left[1 + \left(1+\alpha\right)\frac{t_{\rm vap}}{t_0}\right]^{\frac{1}{1+\alpha}}\;,
\label{tvap_in}
\end{equation}
where $T_{\rm vir}$ is the initial virial temperature of the first core as
given by equation \ref{tvir_core}. As an example, in the limiting cases when
$t_{\rm vap} \gg t_0$, the results are:
\begin{equation}
t_{\rm vap} = 1.3 \times 10^4 \; \hbox{years}\; \left(\frac{T_{\rm vap}}{1200\,\hbox{K}}\right)^2 m_1^{-2} T_1^{-2/9} \kappa_{*}\;,
\label{t2nd1}
\end{equation}
for $\alpha = 1$, and 
\begin{equation}
t_{\rm vap} = 1.6 \times 10^6 \; \hbox{years}\; \left(\frac{T_{\rm
    vap}}{1200\,\hbox{K}}\right)^3 m_1^{-2} T_1^{8/9} \kappa_{*}\;
\label{t2nd2}
\end{equation}
for $\alpha = 2$. These time scales are very long, and are strongly dependent
on the mass of the first core. The term neglected in the last two equations
becomes important for more massive cores that are already hot at their
birth. They may reach the vaporisation temperature very quickly. 

\begin{figure}
\centerline{\psfig{file=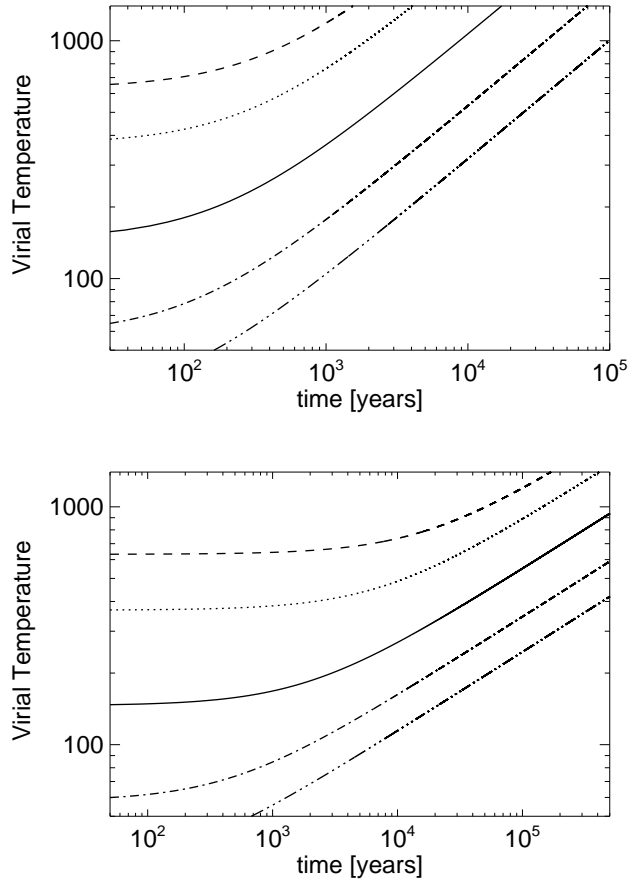,width=0.5\textwidth,angle=0}}
\caption{Virial temperature as a function of time in the analytical model for
  cores of different masses, $\mfc = 3, 5, 10, 20$ and 30 ~$M_J$, from bottom
  to top curves, respectively. The {\bf top} panel is for opacity power-law
  index $\alpha=1$, whereas the {\bf bottom} panel is for $\alpha = 2$. The
  plots are terminated at $T =1400$ K, when even the largest grains would
  vaporise rapidly. Note that lower mass cores could support grain growth for
  longer as they are initially cooler.}
\label{fig:tvir}
\end{figure}

Figure \ref{fig:tvir} shows the evolution of the virial temperature (simply
referred to as temperature of the first cores hereafter) for several values of
$\mfc$ as a function of time for the two limiting opacity cases. The curves
are computed using equation \ref{Toft}. The upper panel presents the faster
contracting case $\alpha=1$ and the lower one corresponds to $\alpha=2$. In
the case of the lower opacity ($\alpha=1$), the relatively massive cores,
$\mfc \simgt 20 M_J$, contract quite rapidly during the first few thousand
years. In terms of grain growth and sedimentation, these cores are not very
promising, as we shall see below. In contrast, for $\mfc \simlt 10 M_J$
for $\alpha=1$, and for all values of $\mfc$ for $\alpha=2$, the cores take $t
> 10^4 - 10^5$ years to contract to the grain vaporisation temperature. Grain
growth is plausible in these cases.

\begin{figure}
\centerline{\psfig{file=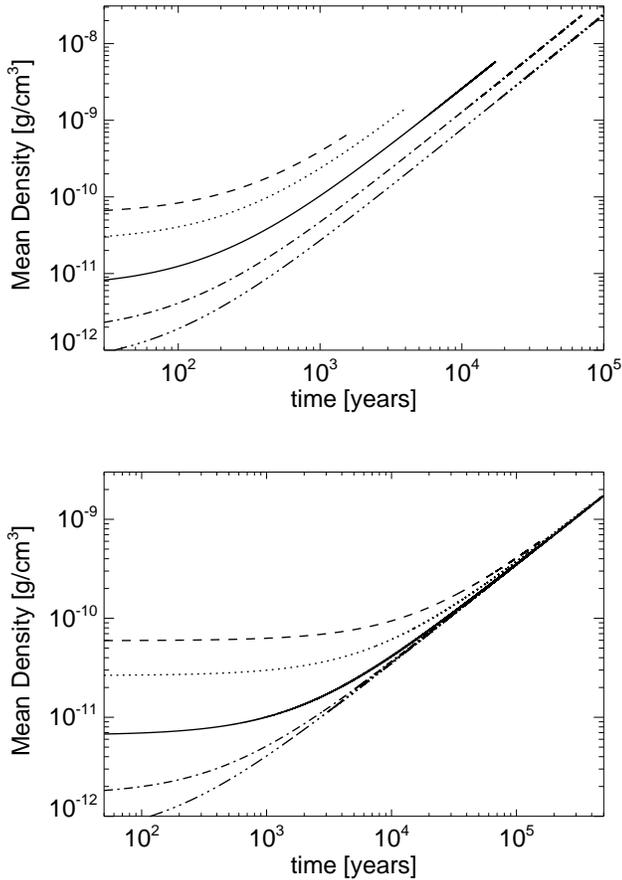,width=0.5\textwidth,angle=0}}
\caption{Same as in Figure \ref{fig:tvir}, but now showing the mean density of
  the first cores. The mean density curves are terminated when first core's
  temperature reaches $T = 1400$ K. Note that maximum mean densities are a few
  $\times 10^{-10}$ g cm$^{-3}$ for the more massive cores but can reach $\sim
  10^{-8}$~g~cm$^{-3}$ for the least massive cores.}
\label{fig:doft}
\end{figure}

We also plot first core's mean densities as a function of time in Figure
\ref{fig:doft}. As in Figure \ref{fig:tvir}, the curves are terminated when
the core's temperature rises above $1400 K$, as then the grains would be
vaporised. We see that the lower mass first cores evolve (contract)
significantly before vaporisation of grains occur. This again hints at less
massive first cores as the more promising sites of grain growth and
sedimentation, as they simply stay cooler for longer.

\subsection{Time scales for grain growth and sedimentation}\label{sec:free_growth1}

\cite{Boss98} studied formation of gaseous giant planets by gravitational
instability in a proto-planetary disc. He considered a gaseous clump that
would eventually contract to form a proto-planet of mass $M=1
M_J$. \cite{Boss98} demonstrated that if the conventional arguments for dust
growth and sedimentation are specialised for the spherical geometry of the
clump, then one can expect a rather rapid dust growth and sedimentation which
may culminate in a formation of a heavy elements core.

Because of the specifics of the gravitational disc instability model for
planet formation, the gas density of $10^{-8}$ g cm$^{-3}$ was used for the
gaseous clump. As shown in \S \ref{sec:1st_analytical}, our gaseous clumps are
much less dense and are also cooler in their initial stages. Nevertheless, the
line of arguments of \cite{Boss98} for the dust growth and sedimentation model
can be simply rescaled to the problem at hand. Our treatment thus follows his model.

We shall first assume that turbulent motions in the first core can be
neglected. This allows us to make the most optimistic estimates of grain
growth and sedimentation. The importance of turbulence will be elaborated in
\S \ref{sec:turbulence}, and further tested with numerical models in \S 3.

Further, we note that realistically, grains of different sizes are present in
the cloud at any given time, due to fragmentation of larger grains by high
speed collisions. Only a fraction of grains is large enough to start
sedimenting down. Therefore the mass fraction $f_g$ in our model should be
thought to represent only those larger grains. A sizable fraction of the
original grains is assumed to remain small and tightly bound to the
gas. Accordingly, a constant gas opacity of the form given by equation
\ref{kappa0} is used despite allowing larger grains to sediment.

Following \cite{Safronov69,Weiden80,Boss98}, we assume that the grains can
grow by the hit-and-stick mechanism \citep[but see][]{BlumWurm08}. For a
constant density core, the gravitational acceleration, $a_{\rm gr}$,
\begin{equation}
a_{\rm gr} = - \frac{4 \pi}{3} G \rho_{\rm fc} R\;,
\label{ag}
\end{equation}
where $R\le \rfc$ is the radial position of the grain inside the first core.

There is no pressure gradient force for the grains, but there is a gas-grain
drag force if the grain velocity, $u_a$, is different from that of gas, $u$.
Note that $u_a$ is always subsonic with respect to the gas speed of sound, as
the largest velocity that the grain can attain is the free-fall velocity which
is of the same order as $c_s$ by definition. 

The drag force on a spherical body depends on the Reynolds number, $Re = a
|\Delta u|/\lambda c_s$, where $\Delta u = u_a - u$, and $\lambda$ is the mean
free path for hydrogen molecules in the gas. The latter is relatively large,
e.g., $\lambda = 1/(n \sigma_{\rm H2}) \approx 40$ cm $\rho_{-10}^{-1}$, where
$\rho_{-10}=10^{10} \rho_{\rm fc}$ is the dimensionless density of the first
core. As we shall see below, grains will usually satisfy $a \simlt \lambda$,
in which case the Epstein drag law applies, and one has \citep{Boss98}
\begin{equation}
\left(\frac{d u_a}{dt}\right)_{\rm drag} = - \frac{\rho_{\rm fc}
  c_s}{\Sigma_a} (u_a-u)\;.
\label{friction_epstein}
\end{equation}
Here $a$, $\Sigma_a = \rho_a a$ and $\rho_a \sim 1$ g cm$^{-3}$ are the
radius, the column and the mass density of the grain, respectively. 

For a larger body, $a > \lambda$, the Stokes drag law applies. The law also
depends on the relative velocity $\Delta u$ through the Reynolds number
\citep[see][]{Weiden77}. In the limit of the small $Re<1$, The Stokes law
yields
\begin{equation}
\left(\frac{d u_a}{dt}\right)_{\rm drag} = - \frac{3\rho_{\rm fc}
  c_s}{2\Sigma_a} \frac{\lambda}{a}\;(u_a-u)\;.
\label{friction_stoke}
\end{equation}

Neglecting the complicated behaviour of the drag coefficient for intermediate
values of $Re$, we combine both regimes described by equations
\ref{friction_epstein} and \ref{friction_stoke} into one, approximately, as
\begin{equation}
\left(\frac{d u_a}{dt}\right)_{\rm drag} = - \frac{\rho_{\rm fc} c_s}{\Sigma_a}
\frac{\lambda}{a+\lambda} (u_a-u)\;.
\label{friction2}
\end{equation}
Given our simple one size dust model, a better gas drag treatment appears to
be excessive, but future more detailed calculations should utilise more
careful drag force treatments.

The equation of motion for the grain is then 
\begin{equation}
\frac{d u_a}{dt}  = - \frac{\rho_{\rm fc} c_s}{\Sigma_a}
\frac{\lambda}{a+\lambda} (u_a-u) - \frac{4 \pi}{3} G \rho_{\rm fc} R\;.
\label{eqm_tot}
\end{equation}
Small grains, $a \ll \rho_{\rm fc} R/\rho_a$, quickly reach their terminal
velocity, so that $du_a/dt \approx 0$, and the grains slip through the gas in
the direction of the centre of the cloud with the sedimentation velocity
\begin{equation}
-u_{\rm sed} = (u_a - u) = - \frac{4 \pi G \Sigma_a R}{3 c_s} \;\frac{\lambda+a}{\lambda}\;.
\label{vsed}
\end{equation}
Note that this velocity is proportional to $R$. For this reason, for a given
radius of the grain $a$, grains starting at different $R$ will fall to the
centre at the same time. We refer to this time scale as the sedimentation or
dust settling time:
\begin{equation}
t_{\rm sed}= \frac{R}{u_{\rm sed}} \approx \frac{3 c_s}{4 \pi G \Sigma_a}
\frac{\lambda}{a+\lambda}\approx 5 \times 10^3 \;\hbox{yrs}\; m_1^{2/3}
(\rho_a a)^{-1}\;,
\label{tseta}
\end{equation}
where the density of the grain $\rho_a$ and its size $a$ are in cgs units, and
we assumed $a\ll \lambda$ limit in the last step.

For microscopic grains, say 1 micron, the sedimentation time scale is
prohibitively long. Therefore grains must become larger before any
sedimentation takes place. The early growth of microscopic grains is 
dominated by Brownian motions of the smallest grains \citep{DD05}. 
Since the sedimentation velocity is a function of the grain
size $a$, grains of different size move with differential speeds. Hence the
larger grains sweep smaller grains, leading to the growth of the grain's mass,
$m_a = (4\pi/3) \rho_a a^3$, at the rate
\begin{equation}
\frac{dm_a}{dt} = \pi a^2 f_g \rho u_{\rm sed}\;,
\end{equation}
where $f_g=0.01$ is the mass fraction of grains, so that $\rho_g = f_g \rho$
is the density of grain material. This translates into the growth rate for $a$
of
\begin{equation}
\frac{da}{dt} = \frac{f_g \rho}{4\rho_a} u_{\rm sed}\;.
\label{dadt1}
\end{equation}
One finds that this differential settling grain growth is faster than that due
to Brownian motion for larger grains.  We now define the grain size e-folding time
scale, $t_{\rm e} = a/(da/dt)$,
\begin{equation}
t_{\rm e}(0) = \frac{3 c_s}{\pi f_g \rho_{\rm fc} G R}\;.
\label{tgr_e}
\end{equation}
Using $R\sim \rfc$, the grain growth time scale is defined as the time needed
to increase the grain size from an initial value $a_0$ to the final size $a$:
\begin{equation}
t_{\rm gr} = \frac{3 c_s}{\pi f_g \rho_{\rm fc} G \rfc}\; \ln\frac{a}{a_0}\;.
\label{tgr}
\end{equation}
Choosing $a = 10$ cm and $a_0 = 10^{-4}$ cm as an example, $\ln (a/a_0)
\approx 23$. Noting that $\rfc/c_s = t_{\rm dyn} \sim (G \rho)^{-1/2}$, we see
that for $f_g = 0.01$, some $2\times 10^3$ dynamical times need to pass before
the grains attain a size at which they can sediment rapidly. Using the
relations from \S \ref{sec:1st_analytical}, we have
\begin{equation}
t_{\rm gr}(0) = 5.4 \times 10^4  m_1^{-1} f_{-2}^{-1} k_{*}^{2/3}
T_1^{-(5+20\alpha)/18} \frac{\ln(a/a_0)}{20}\;,
\label{tgr_numbers}
\end{equation}
in years, where $f_{-2} = f_g/0.01$. In this equation we emphasised the fact
that the grain growth time estimate is obtained for $t=0$.

\subsection{Time-dependent grain growth}\label{sec:time_dep_growth}

Comparing the grain growth time scale with the cooling times of the first
cores (equations \ref{tcool1} and \ref{tcool2}), and with the longer
vaporisation times (equations \ref{t2nd1} and \ref{t2nd2}), one is tempted to
think that grain growth is too slow to be of importance, especially for the
lower opacity index, $\alpha=1$. However one needs to be more careful here as
the first core contracts with time. Therefore grain growth actually
accelerates as the gas cloud cools, and especially rapidly for the case
$\alpha=1$.

\begin{figure}
\centerline{\psfig{file=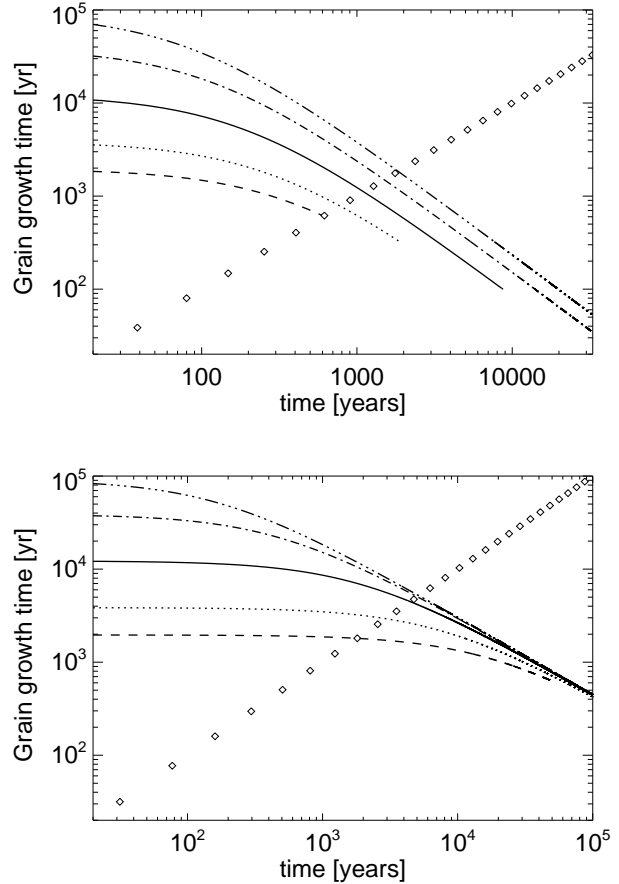,width=0.5\textwidth,angle=0}}
\caption{Grain growth time scale, $t_{\rm gr}$ (equation \ref{tgr}), for
  time-evolving first cores of different masses and the two opacity cases
  (cf. Figures \ref{fig:tvir} and \ref{fig:doft}). The diamonds show the line
  $t_{\rm gr}=t$. Anywhere where the $t_{\rm gr}$ curve drops below the
  diamond line, the grain growth is possible. The curves are terminated when the
  cores become hot enough to vaporise the grains. Lower opacity ($\alpha=1$)
  and higher mass first cores cannot support substantial grain growth and
  sedimentation.}
\label{fig:tgr}
\end{figure}

Using equation \ref{tgr} and our model for the evolution of the first cores,
we plot the time-dependent estimate for the grain growth time scale in Figure
\ref{fig:tgr} for the same first core masses as in Figures \ref{fig:tvir} and
\ref{fig:doft}. Although the lower mass cases (upper curves in Figure
\ref{fig:tgr}) initially have longer grain growth time scales, they are
also initially cooler and stay sufficiently cool to avoid the second collapse for
longer. Therefore, these cores should be promising sites of grain growth after
a few hundred years, when they cross the line shown with diamonds, which is
simply the current time.

We can improve the analytical estimate for the time needed for grain growth
(equation \ref{tgr_numbers}) by utilising the results from \S
\ref{sec:evolution}. We observe that $t_{\rm gr} \propto c_s/(\rho_{\rm fc}
\rfc) \propto (\mfc/\rfc)^{1/2} (\rfc^2/\mfc) \propto \tilde T(t)^{-3/2}$ at a constant
$\mfc$. Therefore, the equation
for the grain growth in a contracting first core is
\begin{equation}
\frac{da}{dt} = \frac{a}{t_{\rm e}(0)} \left[1 +
  \left(1+\alpha\right)\frac{t}{t_0}\right]^{\frac{1.5}{1+\alpha}}\;.
\label{dadt_time}
\end{equation}
This equation can be integrated analytically, and inverted to give a more
accurate estimate for the grain growth time from size $a_0$ to $a$:
\begin{equation}
t_{\rm gr} = \frac{t_0}{1+\alpha}\;\left[  \left(1 +
  \frac{2.5+\alpha}{t_0}t_{\rm gr}(0) \right)^\xi -1
\right]\;,
\label{tgr_exact}
\end{equation}
where $\xi = (1+\alpha)/(2.5 + \alpha)$. In practically all the interesting
parameter space, $t_{\rm gr}(0)\gg t_0$, and the above equation can be further
simplified to
\begin{equation}
t_{\rm gr} = \frac{(2.5+\alpha)^\xi}{1+\alpha} t_0^{1-\xi} t_{\rm gr}(0)^\xi\;.
\label{tgr_simple}
\end{equation}
This results in considerably shorter grain growth times. In particular, for
$\alpha=1$, omitting the dependence on $T_1$,
\begin{equation}
t_{\rm gr}(a) = 3.3 \times 10^3 m_1^{-2/7}f_{-2}^{-4/7} k_{*}^{9/21} \left(\frac{\ln(a/a_0)}{20}\right)^{4/7}\;,
\label{tgra1}
\end{equation}
and for $\alpha=2$,
\begin{equation}
t_{\rm gr}(a) = 2.3 \times 10^4 f_{-2}^{-2/3} k_{*}^{1/3} \left(\frac{\ln(a/a_0)}{20}\right)^{2/3}\;.
\label{tgra2}
\end{equation}
Both of these cases show that grain growth is possible within a reasonable
range of parameters, as the vaporisation time $t_{\rm vap}$ can be comfortably
longer than this. Furthermore, numerical experiments (\S
\ref{sec:num_isolated}) show that grain growth time is about a factor of two
shorter than the analytical model predicts.

\subsection{Sedimentation for a constant grain size}

Based on our numerical experiments below, grains do not sediment strongly
until their size reaches at least a few cm.  Approximately then, we divide the
process of grain growth and sedimentation into the growth phase, studied
above, and the sedimentation phase. Assuming that grain growth slows down in
the latter phase (due, e.g., to a too high relative velocity between the
grains, see \S \ref{sec:destr_coll}), we consider the grains to have a
constant size whilst they sediment. This simplified model allows analytical
calculations to be carried through. We shall continue to approximate the first
core with the constant density $\rho = \rho_{\rm fc}$ for $R \leq \rfc$ and
zero density outside, as in \S \ref{sec:1st_analytical}. Furthermore, we
shall fix these, essentially assuming that sedimentation occurs on a time
scale shorter than the current cooling time of the first core.

\subsubsection{Homologous contraction in the gas-dominated
  phase}\label{sec:homology}

We start off with the grain density radial distribution following the gas
distribution, thus $\rho_g(R, 0) = f_g \rho_{\rm fc}$, and zero grain
velocity. The gas is assumed stationary. One can show that the grains achieve
the terminal velocity $u = - u_{\rm sed}$ (equation \ref{vsed}) very quickly
when $|v_a| \ll c_s$. Here we kept the dependence of the drag force on the
grain size in the more general form (equation \ref{friction2}). Note that the
sedimentation time scale $t_{\rm sed}$ is independent of $R$ in the limit
$a\ll \lambda$ (see equation \ref{tseta}).
Radial position of grains evolves according to 
\begin{equation}
\frac{d R}{dt} = - \frac{R}{t_{\rm sed}}\;,
\end{equation}
which is trivially solved,
\begin{equation}
R(t) = R_0 \exp\left[-\frac{t}{t_{\rm sed}}\right]\;,
\label{rsed}
\end{equation}
where $R_0$ is the initial value of $R$.  $R(t)$ is the Lagrangian coordinate
for a grain mass shell $M_g(R_0)$. This equation shows that the grains
contract homologously, so that the grain density is independent of radius
within $R_{\rm g}(t) = \rfc \exp(-t/t_{\rm sed})$, and is zero outside.  The
grain density profile keeps its top-hat shape but becomes more compact. The
grain density increases with time as
\begin{equation}
\rho_g(t) = f_g \rho_{\rm fc} \exp\left[\frac{3t}{t_{\rm sed}}\right]\;.
\label{rhosed}
\end{equation}

\subsubsection{Grain-dominated core phase}\label{sec:grain_dominated} 

The equation \ref{vsed} for the grain settling velocity neglects gravity from
grains, which is appropriate in the initial stages of the process. When the
density of the grains approaches and then exceeds that of the gas, the
equation obviously becomes inaccurate.  The contracting cloud of grains
becomes self-gravitating at the time
\begin{equation}
t_{\rm self} = t_{\rm sed} \frac{\ln (f_g^{-1})}{3} = 1.53 \; t_{\rm sed}\;.
\label{tself}
\end{equation}
The grain sphere's outermost radius at that time is
\begin{equation}
R_{\rm self} = \rfc f_g^{1/3} \approx 0.2 \left[\frac{f_g}{0.01}\right]^{1/3} \rfc\;.
\label{radsed}
\end{equation}
Thus to follow the contraction of the grain sphere after time $t_{\rm
  self}$, we modify the equation for settling velocity by writing
\begin{equation}
\frac{dR}{dt} = - \frac{G M_g(R_0)}{R^2} \frac{\Sigma_a (a +
  \lambda)}{\rho_{\rm fc} \lambda c_s}\;,
\label{vcontr}
\end{equation}
which now completely neglects the mass of the gas interior to radius $R$, as
$\rho_g \gg \rho_{\rm fc}$, asymptotically. Since $M_g(R) = M_g(R_0)$, i.e.,
constant in Lagrangian coordinates, the above equation is solved as
\begin{equation}
R(t) = R(t_{\rm self}) \left[1 - 3\,\frac{t-t_{\rm self}}{t_{\rm sed}}\right]^{1/3}\;,
\end{equation}
which is a homologous contraction again, albeit at a different -- accelerated
by the self-gravity -- rate. The density evolution follows the form
\begin{equation}
\rho_g(t) = \rho_{\rm fc}  \left[1 - 3\,\frac{t-t_{\rm self}}{t_{\rm
      sed}}\right]^{-1}\;,
\label{rhogt1}
\end{equation}
where we utilised the fact that the grain density at the initial time when the
solution \ref{rhogt1} becomes applicable, $t=t_{\rm sed}$ is equal to
$\rho_{\rm fc}$.  Within this simple model, all the grains collect to the
centre of the first core, reaching formally infinite densities, at time
\begin{equation}
t_{\infty} = t_{\rm self} + \frac{t_{\rm sed}}{3} = \frac{\ln
  (e/f_g)}{3}\;t_{\rm sed} = 1.87 \; t_{\rm sed}\;
\label{tinf}
\end{equation}
for $f_g = 0.01$.

\subsubsection{Bound ``grain cluster'' phase}\label{sec:grain_cluster}

Before contraction of the grains into a point occurs, another important
milestone is reached when the grain sphere becomes not only self-gravitating
but also gravitationally self-bound. In the self-contracting phase studied
above, the gravitational force acting on the grains is dominated by the grain
density. Hence the grains are self-gravitating in that sense. However, if the
outer gaseous envelope were removed in the beginning of that phase, gas inside
the grain sphere would create a significant pressure gradient that could
unbind the contracting grain-gas mix. In contrast, at later time, when the
grain sphere contracts even further, its density is high enough that removal
of the gaseous envelope would not unbind the grains. We refer to this stage as
the ``grain cluster'' one, in analogy to a star cluster.

Consider then the question of how compact the grain sphere of mass $f_g \mfc$
should be in order to be self-bound, i.e., so that the gas pressure could not
unbind the sphere. We follow the usual order of magnitude arguments with which
the Jean's mass can be derived. The gas pressure gradient is $\sim \rho k_B
T/\mu R$, whereas the gravitational acceleration is $(GM_{\rm enc}(R)/R^2)
(\rho + \rho_g)$. The total enclosed mass is $M_{\rm enc}(R) = (4\pi/3)
(\rho + \rho_g) R^3$. This defines the radial scale (Jean's length and also
the size of the grain cluster)
\begin{equation}
R_{\rm gc} = \left[ \frac{3}{4\pi} \frac{k_B T}{\mu G} {\rho \over \rho_g^2}\right]^{1/2}\;,
\label{rgc}
\end{equation}
where we explicitly assumed $\rho \ll \rho_g$. The Jeans mass is equal to $M_J
= M_{\rm gc} = (4\pi/3) \rho_g R_{\rm gc}^3$, which can be re-written as
\begin{equation}
M_{\rm gc} \approx \left(\frac{3}{4\pi\rho}\right)^{1/2} \left(\frac{k_B T}{\mu
  G}\right)^{3/2}  \left({\rho \over \rho_g}\right)^{2} \;.
\label{mgccc}
\end{equation}
The first two factors on the right hand side of the equation \ref{mgccc} is
the mass of the first core, $\mfc$. Further,  $M_{\rm gc} = f_g \mfc$, and
hence we can solve for the grain density in the grain cluster:
\begin{equation}
\rho_{\rm gc} \approx \rho_{\rm fc} \; f_g^{-1/2}\;.
\label{rho_gc}
\end{equation}
Using this result in equation \ref{rgc}, the radius of the grain cluster is
found to be
\begin{equation}
R_{\rm gc} \approx \rfc \; f_g^{1/2}= 0.1 \rfc \left[\frac{f_g}{0.01}\right]^{1/2}\;.
\label{rgc2}
\end{equation}
This equation shows that the grain sphere becomes self-bound when it contracts
to size $0.1 \rfc$, e.g., $0.1-1$ AU for a realistic range on parameters. 

Note that the grain density in the grain cluster is typically a factor of ten
or more higher than the density of the first core itself, i.e., as high as
$10^{-9}-10^{-7}$ g cm$^{-3}$.

Presumably, once the ``grain cluster'' phase is reached, nothing keeps the
grain-dominated region from a self-gravitational collapse, so that a solid
core could be formed \citep{Boss98}. We delay the study of this issue till a
future paper.

\subsection{Turbulent mixing}\label{sec:turbulence}

Turbulent mixing \citep{FromangPap06} is a process in which grains are dragged
along turbulent motions of the gas. As initially grains represent a small
fraction of the total mass, they are essentially a trace population, and
therefore gas turbulence drives diffusion of grains.  This process hence tends
to erase grain density inhomogeneities, opposing gravitational settling of
the grains.  

Numerical simulations (\cite{FromangPap06}, and references thereafter) show
that a simple diffusion equation approach describes the effects of turbulence
on dust well, provided that the diffusion coefficient, $D$, is chosen
right. The diffusion equation for grains in the spherical geometry can be
written as
\begin{equation}
\frac{\partial \rho_g}{\partial t} = \frac{D}{R^2} \frac{\partial}{\partial
  R}\left[R^2 \rho \frac{\partial (\rho_g/\rho)}{\partial R} \right]\;,
\label{diff_eqn}
\end{equation}
where $\rho$ and $\rho_g$ are the gas and the grain (dust) density,
respectively. Note that we assumed that $D$ is a constant inside the giant
embryo, e.g., independent of $R$. The diffusion coefficient in case of disc
can be parameterised in the form $D = \alpha_d c_s H$, reminiscent of the
standard disc viscosity prescription \citep{Shakura73}, where $\alpha_d < 1$
is the viscosity coefficient, and $H$ is the disc vertical scale height. We
shall use the same approach, except that the scale-height $H$ should be
replaced by the radius of the first core, $\rfc$:
\begin{equation}
D = \alpha_d c_s \rfc\;.
\label{diff_coeff}
\end{equation}
The diffusion coefficient $\alpha_d$ is unknown here.  In the disc
geometry, differential rotation velocity develops between the dust layer and
the gas, driving instabilities \citep[see, e.g.,][]{GaraudLin04}. The
simulations of the turbulent disc by \cite{FromangPap06} showed that $\alpha_d
\sim 0.004$, but there is no clear reason why their results should translate
to the case we study. Therefore we shall treat $\alpha_d$ as a free parameter
of the model and consider the implications of turbulent diffusion in the
``small'' and ``large'' $\alpha_d$ cases.

To continue with our analytical modelling for now, we shall again assume the
``top hat'' density profile for the gas, and hence set $\rho$ to a
constant. Also, we shall combine the diffusion equation for dust (grains) with
the mass continuity equation:
\begin{equation}
\frac{\partial \rho_g}{\partial t} = \frac{D}{R^2} \frac{\partial}{\partial
  R}\left[R^2 \frac{\partial \rho_g}{\partial R} \right] -  \frac{1}{R^2} \frac{\partial}{\partial
  R}\left[R^2 \rho_g V \right]\;,
\label{drhodt_full}
\end{equation}
where $V = - u_{\rm sed}$ is the grain terminal velocity derived earlier
  (equation \ref{vsed}).

In a steady state, the solution of this equation implies
\begin{equation}
\rho_{\rm eq} = \rho_{d0} \exp\left[ -\frac{R^2}{H_d^2} \right]\;,
\label{rho_steady}
\end{equation}
where $H_d$ is the ``dust sphere'' scale height, given by
\begin{equation}
H_d = \left(2 D t_{\rm sed}\right)^{1/2}
\label{hdeq}
\end{equation}
Using equation \ref{tseta} for sedimentation time $t_{\rm sed}$ and $c_s^2 =
G\mfc/\rfc$, we obtain
\begin{equation}
\frac{H_d}{\rfc} = \left[ \frac{3\alpha_d}{2} \frac{\Sigma_{\rm fc}}{\Sigma_a}
  \frac{\lambda}{\lambda +a} \right]^{1/2}
\label{hdeq_tor}
\end{equation}
where $\Sigma_{\rm fc} = \mfc/(\pi \rfc^2) \sim 10^3$ g cm$^{-2}$ is the
column depth of gas in the first core (equation \ref{sigma_core}). Recalling
that $\Sigma_a = \rho_a a$, we see that for any grain sedimentation to take
place (i.e., to have $H_d \ll \rfc$), we need $\alpha_d$ to be small and $a$
to be large, e.g., macroscopic. For example, if $\alpha_d = 0.001$, $a <
\lambda $, and $\rho_a\sim 1$ g cm$^{-3}$, grains larger than 1 cm are
required.

\subsection{Destructive collisions}\label{sec:destr_coll}

Experiments show that the simple hit-and-stick picture for grain growth is
modified when the relative velocity is larger than a few metres per second
\citep{BlumWurm08}. Above these speeds grains can fragment or stick only
partially. Further growth-reducing processes such as cratering or partial
fragmentation can occur. To estimate the potential importance of this, 
we recall that the terminal velocity of a grain of size $a$ is
\begin{equation}
u_{\rm sed} = \frac{4 \pi G \Sigma_a R}{3 c_s} = 4.4 \;\hbox{m s}^{-1}\; a
\;\frac{R}{\rfc} \tilde T(t)^{-3/2}\;,
\label{vset10}
\end{equation}
where $a\ll \lambda$ is in cm, and we assumed $\mfc = 0.01 \msun$. Here the
time-dependent dimensionless function $\tilde T(t)$ (equation \ref{Toft})
accounts for the fact that the first cores contract with time, so that
$\rfc/c_s \propto \rho_{\rm fc}(t)^{-1/2} \propto T(t)^{-3/2}$.

From the above equation we see that growth of cm-sized grains may stall as
their sedimentation velocity exceeds several metres per second. For a
qualitative estimate of the effects of this, we can further assume that grain
growth saturates at a size $a$ such that $u_{\rm sed} = v_{\rm max}$, where
$v_{\rm max}$ is a few metres per second. In this case the grain sedimentation
time scale is modified to
\begin{equation}
t_{\rm fragm} = \rfc/v_{\rm max} = 3 \times 10^4\;\hbox{yr}\;
\frac{\rfc}{3\times 10^{13}}\frac{3\,\hbox{m s}^{-1}}{v_{\rm max}}\;.
\label{tfragm}
\end{equation}
This ``fragmentation-limited'' sedimentation is slow, but not prohibitively
slow compared with the grain growth time and the vaporisation times calculated
earlier. The qualitative indication is hence that fragmentation of grains may
slow down but not stop grain sedimentation inside the first cores.

\section{Set up of numerical models}\label{sec:num_isolated}

Having learned about the problem perhaps as much as possible within a simple
analytical approach, we now turn to a spherically symmetric radiation
hydrodynamics code to follow the evolution of the first cores and grain
sedimentation inside them.  We use Lagrangian (gas) mass coordinates and the
classical radiative diffusion approximation, which is appropriate given that
first cores are very optically thick except for their atmospheres, which are
of little interest to us here.

\subsection{No grain sedimentation case: gas only equations}\label{sec:nogrowth}

\begin{figure}
\centerline{\psfig{file=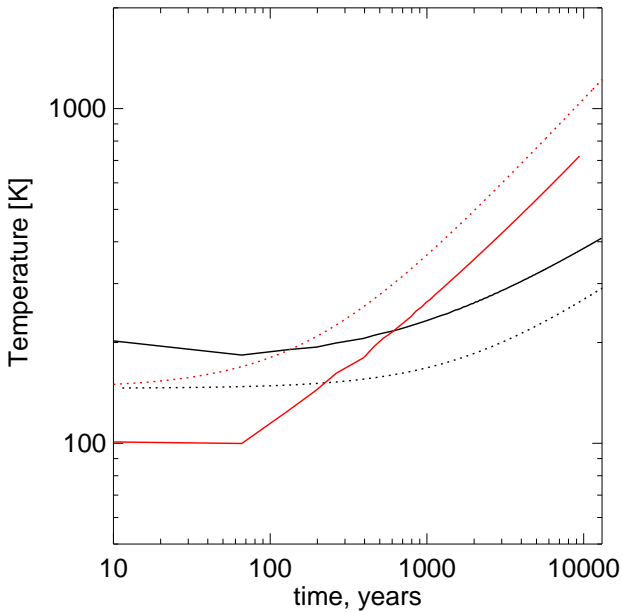,width=0.5\textwidth,angle=0}}
\caption{Mean temperature (solid curves) of the first core of mass $\mfc =
  10 M_J$ as a function of
  time for $\alpha=1$ (red) and $\alpha=2$ (black). The dotted curves show the
  virial temperature of the first cores in the analytical model presented in
  \S \ref{sec:evolution} for same cases. Note that the analytical curves are
  within $\sim 20-30$\% of the numerical calculation.}
\label{fig:tmean}
\end{figure}

First of all we shall present evolution of the first core neglecting dust
growth and sedimentation. Besides being interesting in itself, such a
calculation is useful to contrast with the analytical solutions we obtained in
\S \ref{sec:evolution}. The equations being solved are the standard
hydrodynamical equation with the addition of the radiative diffusion cooling:
\begin{eqnarray}
\frac{d u}{dt} = - 4\pi R^2 \frac{d P}{dM} - \frac{GM_{\rm tot}(R)}{R^2} +
a_{\rm g-d}\;,\\
\label{rad_f}
\frac{d \varepsilon}{dt} = - 4\pi P \frac{d (R^2 u)}{dM} - 4\pi \frac{d (R^2
  F_{\rm rad})}{dM}\;,\\
\frac{d R}{dt} = u\;,
\end{eqnarray}
where $u$ is gas velocity, $M_{\rm tot}(R) = M(R) + M_g(R)$ is the total (gas
+ grains) mass enclosed within $R$, $a_{\rm g-d}$ is the acceleration due
to the gas-dust drag, to be discussed below, $P$ is gas pressure, $F_{\rm
  rad}$ is the classical radiation flux, given by
\begin{equation}
F_{\rm rad} = - \frac{4 a_{\rm rad} cT^3 }{3\kappa(T)}\,4\pi R^2\,
\frac{dT}{dM}\;,
\label{flux}
\end{equation}
where $a_{\rm rad}$ is the radiation constant. The gas density is defined by
$\rho^{-1} = 4\pi R^2 dR/dM$. The numerical time integration procedure is
based on the Lagrangian scheme ``lh1'' presented in
\cite{BodenheimerBook}. The scheme uses artificial viscosity to capture
shocks. We initialise the first cores as polytropic spheres of a given gas
mass $\mfc$ as described in \S \ref{sec:1st_analytical}.

Most of the simulations here do not include the convective energy transfer
since we did not expect it to be important: the optical depth of the first
cores is not very high, so radiative cooling is quite efficient until the gas
clump contracts significantly. This was confirmed in some of the simulations
that did include the convective energy transfer. Furthermore, all of the
simulations in paper II include the convective flux, and also the resulting
convective mixing of the grains. We found that convection is never dominant
before the collapse of the central grain concentration into a massive solid
core. Once that occurs, the release of gravitational energy from the core
significantly changes the entropy profile near the solid core. This change
drives very strong, in fact supersonic, convective motions in the region near
the core. This strongly affects grain dynamics but {\em after} the solid core
has already formed.

The boundary conditions at $R=0$ are $u=0$, $M(0) = 0$, $dP/dM=0$ and $F_{\rm
  rad}=0$. At the outer boundary we require continuity of gas density,
velocity and the radiation flux. In this section we consider the case of dust
tightly bound to the gas, in which case one can simply neglect the dust,
setting $a_{\rm g-d}$ to zero.

Figure \ref{fig:tmean} shows the mean temperature in the cloud as a function
of time for the first core of mass $\mfc=0.01 \mfc$ for the two opacity cases,
$\alpha=1$ and $\alpha=2$. The dimensionless opacity coefficient $k_{*}=1$
and the ambient gas temperature $T_1 =1$ for all the curves in the figure.
The slight jumps in the solid curves are due to initial oscillations of the
first cores, which quickly decay away due to artificial viscosity (the
numerical representation of the polytropic sphere initial conditions results
in small perturbations).  We also plot the analytical solutions (dotted lines)
for the same clouds and opacity cases as obtained in \S
\ref{sec:evolution}. Notably, the analytical and numerical solutions do not
deviate from each other by more than a few tens of percent, which is more than
can be expected from the order of magnitude analytical approach.

\subsection{Inclusion of grain physics}

\subsubsection{Grain dynamics}\label{sec:gr_dyn}

We now turn our attention to the more complex case in which grain dynamics is
taken into account.  We use grains of the same size $a$ for the whole cloud,
where $a$ is a free parameter. A more complex treatment is possible but is
delayed until a future paper.

The grains are treated as a second fluid. We use the Lagrangian radial mesh
provided by the gas coordinate to follow the evolution of dust. This is
natural especially for small size grains that are strongly bound to the gas,
as the relative gas-grain velocity is small, and the grains essentially move
with the gas. The dust is described by a density in a given radial gas shell,
and is allowed to slip through gas shells (see below).  The grain initial
density distribution follows that of gas, scaled down by the factor $f_g$.

The drag force acting on the grains/dust due to friction with the gas,
$(du_a/dt)_{\rm drag}$, is given by equation \ref{friction2}. Due to Newton's
second law, the back reaction on the gas is 
\begin{equation}
\rho \left(\frac{d u}{dt}\right)_{\rm drag} = - \rho_g \left(\frac{d
  u_a}{dt}\right)_{\rm drag}\;,
\label{gas_frict}
\end{equation}
where $u_a$ is grain velocity, $\rho_g$ is dust density (not to be confused
with grain material density, $\rho_a$), and $u$ is gas velocity. Accordingly,
the acceleration $a_{\rm g-d}$ in equation \ref{rad_f} is simply $(du/dt)_{\rm
  drag}$.

The grains themselves are under the influence of gravity, drag force from the
gas, and also undergo turbulent diffusion. From equation \ref{drhodt_full},
this can be accounted for by setting the grain velocity to
\begin{equation}
u_a = u - u_{\rm sed}' - D \frac{\partial \ln(\rho_g/\rho)}{\partial R}
\label{vr_dust_full}
\end{equation}
where $u_{\rm sed}'$ is the grain sedimentation velocity, and the last term is
the grain velocity generated by turbulent mixing. The terminal velocity
approach that is used to derive equation \ref{vr_dust_full} is sufficiently
accurate for small grains. For larger grains, the sedimentation velocity
(equation \ref{vsed}) may be larger than the local free fall velocity. To
correct for this, we use
\begin{equation}
u_{\rm sed}'= \min\left[ u_{\rm sed}, u_{\rm ff}\right]\;,
\label{vsed_prime}
\end{equation}
where $u_{\rm ff} = \sqrt{2GM_{\rm tot}(R)/r}$ is the local free fall
velocity. We tested this approach against the exact integration of grain
radial time-dependent equation of motion, i.e., not assuming a terminal
velocity. The differences are minor, and the terminal velocity approach is more
numerically stable for smaller grains and does not require excessively small
time steps. Therefore we pick the equation \ref{vr_dust_full} as the superior
approach for grain dynamics simulations here.

Note that mass $M_g(R)$, initially exactly equal to $f_g M(R)$, evolves with
time separately from $M(R)$, as grains are allowed to slip through the gas,
and from one gas radial mass shell into another. The mass continuity equation
for grains inside a gas mass shell of index $i$, with inner and outer radii of
$R_i$ and $R_{i+1}$, respectively, is
\begin{equation}
\frac{1}{4\pi} \frac{\partial \Delta M_{g, i}}{\partial t} =
 - \left[R^2 \rho_g (u_{a} - u)\right]_{i+1} + \left[R^2 \rho_{g} (u_{a} - u)\right]_{i}\;,
\label{dmgr_cont}
\end{equation}
where indeces $i+1$ and $i$ refer to the inner and outer boundary of the zone;
e.g., $\rho_{g, i}$ is density of dust in zone $i$, $\Delta M_{g, i} = (4 \pi/3)
\rho_{d, i}( R_{i+1}^3 - R_i^3)$ is the grain mass inside zone $i$.

There is a slight uncertainty in choosing the boundary conditions for grains
at $R\rightarrow 0$ radius, that is in the very first {\em gas} mass zone. In
contrast to the gaseous component, grains are not supported by pressure
effects there, thus they can sediment and form a phase {\em below} even the
very first mass zone of the gas. On the other hand, turbulence may suspend the
grains in the fluid, curtailing the sedimentation into $R=0$. Our goal here is
to be conservative in our calculations of grain sedimentation.  Therefore we
assume that the turbulence keeps the grains suspended in the very first gas
zone {\em as long as} the gas dominates the grains there by mass.
Accordingly, the boundary condition for grain velocity is set to $u_a(R=0) =
0$, as for the gas, for all the numerical tests below. The simulations are
stopped when the grain density in the inner zone reaches that of the gas
density. One would expect that when the grain density exceeds that of the gas,
the feedback that the grains impose on the turbulent motions of the gas
becomes substantial, and hence the turbulent mixing support of grains should
ease off.  Presumably a high density core composed of heavy elements is formed
at this point \citep{Boss98}. A special treatment for the inner boundary
condition is needed when this happens. We defer a study of the core
formation process until a future publication.

\subsubsection{Grain growth and vaporisation}\label{sec:num_ggrowth}

We continue to consider only one size for the grains here, but in the interest
of adding more realism to our models we allow that size to vary -- grow by
sticking with other grains, vaporise if gas enveloping the grain is too hot,
and limit the grain growth if the differential grain velocity is too large.

Our grain growth model follows the analytical prescription of \S
\ref{sec:free_growth1}, but with the addition of Brownian motions of small
grains:
\begin{equation}
\frac{da}{dt} = \left\langle\frac{\rho_g}{4\rho_a} \left(u_{\rm sed} + u_{\rm
  br}\right)\right\rangle\;,
\label{dadt2}
\end{equation}
where $u_{\rm sed}$ is the sedimentation velocity, and $u_{\rm br}$ is the
Brownian motion velocity of the smallest grains that dominate grain relative
velocities at small values of $u_{\rm sed}$ \citep{DD05}. The Brownian motion
velocity is a strong function of the smallest grains size, and we treat it as
a free parameter of the model. The signs $\langle$ and $\rangle$ signify
grain-mass averaging of the quantities over the first core. This approach is a
necessity given that the full radial and particle size distribution function
treatment is beyond the scope of our initial study.

Grain fragmentation or cratering may result from grain collisions with
inter-grain velocities exceeding a characteristic value, which depends on the
size of the grains participating in the interactions \citep{BlumWurm08}. We
explore these effects in a very simple fashion, introducing ``maximum
velocity'', $v_{\rm max}$ , above which grain growth is significantly reduced
compared with equation \ref{dadt2}. In particular, we modify the above
equation to
\begin{equation}
\frac{da}{dt} = \left\langle\frac{\rho_g}{4\rho_a} \left(u_{\rm sed} + u_{\rm
  br}\right)\right\rangle\;   \left(\frac{v_{\rm max}}{v_{\rm max} + u_{\rm sed}}\right)^2 \;.
\label{dadt3}
\end{equation}
With this prescription, the $da/dt \rightarrow 0$ as $u_{\rm sed} \rightarrow
\infty$.

Finally, the grains are allowed to be vaporised if the gas temperature is
large. The vaporisation rate is taken from \cite{HS08}.  In practice this
prescription implies vaporisation temperature between T = 1200 K and T =1500
K, depending on grain size and relevant time scales. Vaporisation is not
important for any models that do not reach 1200 K.


\begin{figure*}
\centerline{\psfig{file=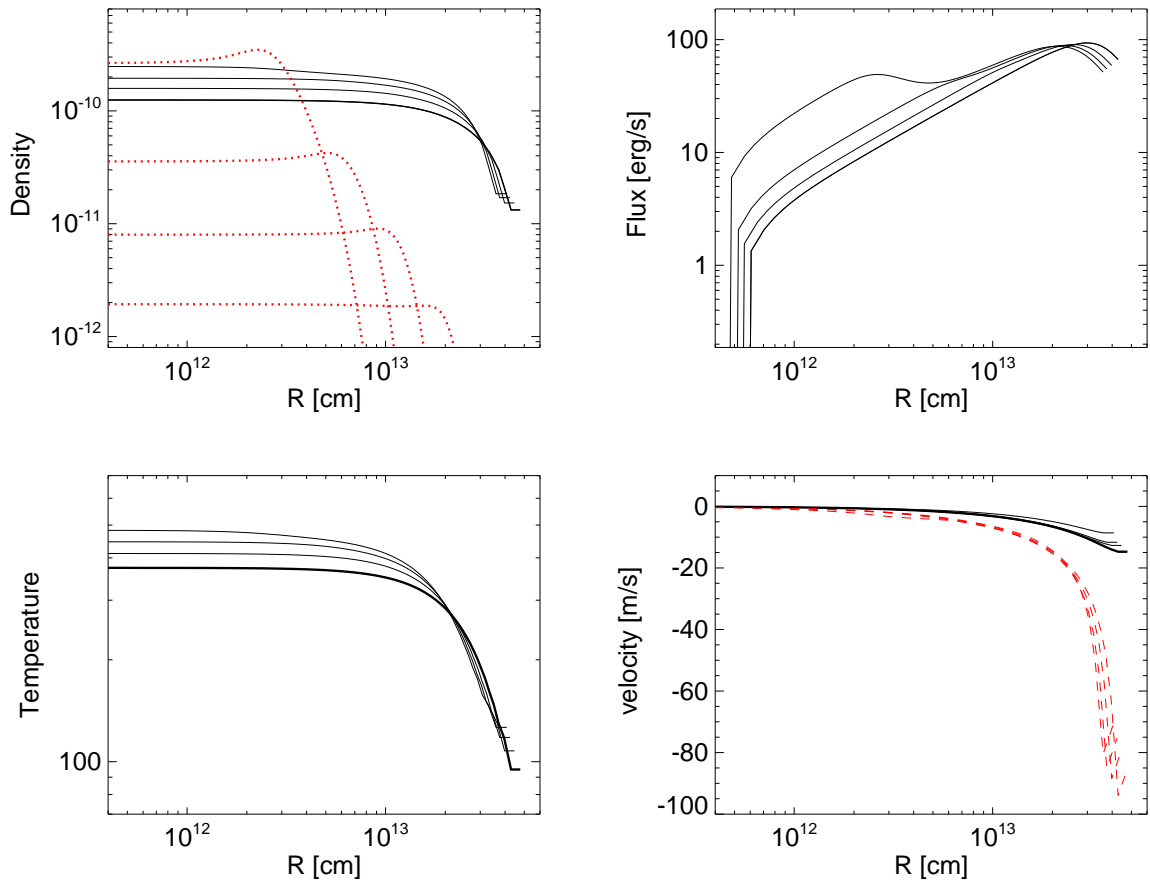,width=0.7\textwidth,angle=90}}
\caption{Grain sedimentation for a first core of mass $\mfc = 10 M_J$ and
  a fixed grain size of $a=10$ cm. The panels show snapshots of gas density,
  temperature, radiation flux and velocity, as labelled, in solid curves. The
  dotted and dashed red curves show the grain density and velocity,
  respectively. The snapshots are for times between $t=360$ and 1500 years.}
\label{fig:fixed_a10}
\end{figure*}

\section{Numerical results}\label{sec:num_results}

\subsection{Constant grain size, no turbulent mixing}\label{sec:free_num}

We start with the simplest, constant grain size case, setting $a=10$ cm. We
neglect the turbulent grain stirring, i.e., fixing $\alpha_d = 0$.  The
calculation assumes the grain material density of $\rho_a = 1$ g cm$^{-3}$,
initial mass fraction of $f_g = 0.005$, opacity coefficient $\kappa_0 = 0.01$,
opacity power-law index $\alpha = 2$, and the first core's mass of $\mfc = 10
M_J$.

Figure \ref{fig:fixed_a10} presents several snapshots showing the evolution of
the gas and the grain components. The snapshots correspond to times $t\approx$
360, 740, 1100 and 1500 years.  The first inference from the Figure is that
the constant density approximations for the first core itself and for the
contracting grain sphere or ``cluster'' are reasonably good.  The analytical
estimate of the settling time for this case is $t_{\rm set} = 500$ years
(equation \ref{tseta}). Further, equation \ref{tself} predicts that the grains
will collect into a central sphere the mass density of which equals that of
the gas at time $t = 880$ years (for $f_g = 0.005$). In the simulations, this
occurs at time almost twice as long. This level of accuracy of the analytical
estimates is nevertheless acceptable to us here. The simulations show a
somewhat more complex evolution than the simple ``top hat'' profile.

The upper right panel of the figure shows the radiation flux as a function of
radius within the first core. Note that in the last snapshot the heat flux
increases significantly in the inner $\sim $ 10\% of the first core. This is a
signature of the adiabatic contraction heating imposed by the ever increasing
grain density in that region. While until this point the effects of grain
sedimentation were hardly felt by the gaseous component anywhere inside the
core, the ``inverse drag'' -- grain drag on gas is now significant in the
inner region. The gas is however stable against further collapse there as can
be seen from velocity curves (lower right panel): the gas and the grain phases
continue to separate out, e.g., move with different velocities.  The reason
for which the gas phase remains gravitationally stable is that it takes a
relatively small amount of heating (the bump in the flux curve in the upper
right panel) to set up a sufficient pressure gradient to oppose the collapse.

The last snapshot of the figure corresponds to a stage between the
grain-dominated one (\S \ref{sec:grain_dominated}) and the ``bound grain
cluster'' (\S \ref{sec:grain_cluster}). Soon after the last snapshot shown in
the figure the grain distribution undergoes a dynamical collapse which we
shall study in a future paper.

\subsection{Constant grain size, turbulent mixing}

\subsubsection{Strong turbulence}\label{sec:fixed_a_strong_turb}

\begin{figure}
\centerline{\psfig{file=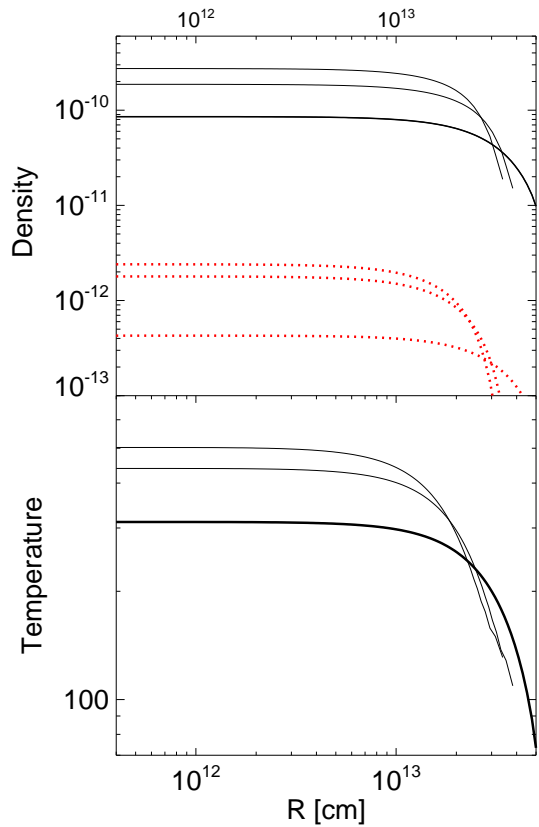,width=0.45\textwidth,angle=0}}
\caption{Density (solid for gas, red dotted for grains) and temperature
  distributions at three different times for the same calculation as in Figure
  \ref{fig:fixed_a10}, but now with the turbulent mixing coefficient $\alpha_d
  = 10^{-2}$. The snapshots are for times $t=0$, 1100 and 2200 years. Note
  that the grain distribution concentrates towards the centre by a small amount
  only before coming to an equilibrium in which sedimentation is balanced by
  turbulent mixing.}
\label{fig:fixed_a10_alpha1m2}
\end{figure}

We now add turbulent mixing, still keeping the size of the grains constant at
$a=10$ cm. Figure \ref{fig:fixed_a10_alpha1m2} shows the density and
temperature distributions for the same case computed in \S \ref{sec:free_num},
but for the turbulent diffusion coefficient $\alpha_d = 0.01$. It is obvious
from the snapshot sequence that the dust distribution quickly adjusts to an
equilibrium shape in which the rate of gravitational settling is offset by
turbulent diffusion. The contraction of the density distribution between the
last two snapshots is entirely due to the contraction of the gas distribution,
which occurs on the radiative cooling time scale of about $10^4$ years. Note
that no dense grain cluster forms in this case no matter how long we were to
follow the calculation. After the first core has contracted enough to reach
$T\sim 2000$ K, it would undergo the well known second collapse
\citep{Larson69,Masunaga00}. Apparently, the effects of grain sedimentation
for these particular parameters are completely negligible.

\subsubsection{Weaker turbulence}\label{sec:fixed_a_mid_turb}

\begin{figure}
\centerline{\psfig{file=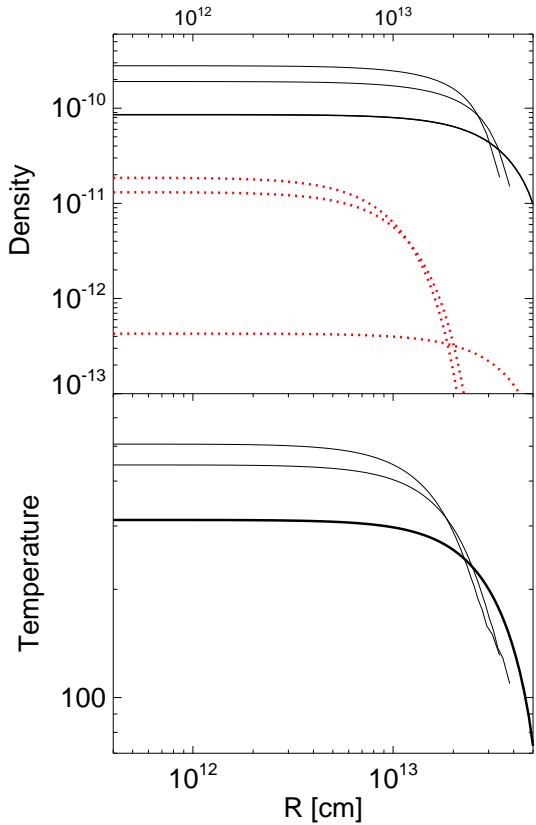,width=0.45\textwidth,angle=0}}
\caption{Same as Figure \ref{fig:fixed_a10_alpha1m2}, but for a smaller
  turbulent mixing coefficient, $\alpha_d = 10^{-3}$. The dust is
  concentrated towards the centre much more than in the previous calculation,
  but still hovers at an equilibrium state.}
\label{fig:fixed_a10_alpha1m3}
\end{figure}

 Figure \ref{fig:fixed_a10_alpha1m3} shows a calculation identical to that
 presented in \S \ref{sec:fixed_a_strong_turb} and in Figure
 \ref{fig:fixed_a10_alpha1m2}, but now for a weaker turbulent mixing, with
 $\alpha_d = 10^{-3}$. Once again an equilibrium is quickly set up, but a much
 more concentrated one. It is interesting to note that this equilibrium
 implies a strong metalicity gradient within the gas cloud, with the most
 metal rich gas being in the central regions.

\subsubsection{Weaker turbulence and larger grains}\label{sec:a30_mid_turb}

\begin{figure}
\centerline{\psfig{file=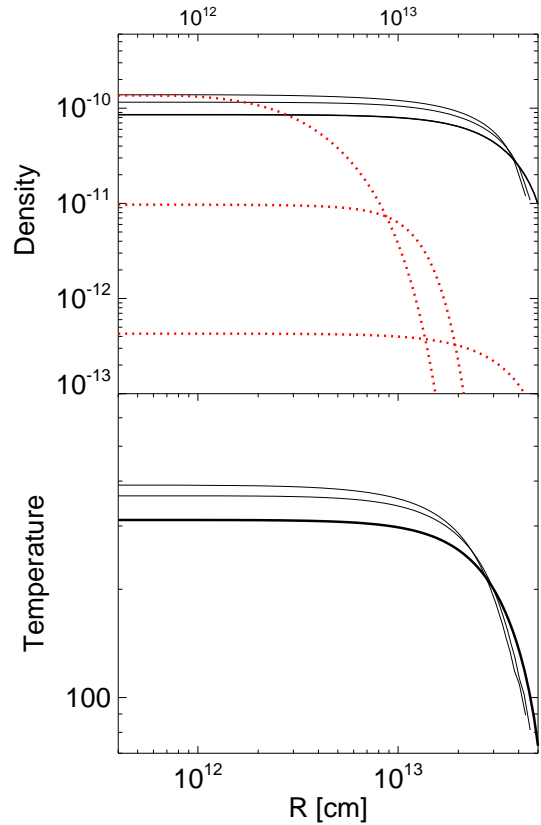,width=0.45\textwidth,angle=0}}
\caption{Same as Figure \ref{fig:fixed_a10_alpha1m3}, but for larger grains,
  $a = 30$ cm. The times of snapshots are $t=$ 0, 220, and 440 years. Soon
  thereafter, the inner grain-dominated part of the first core undergoes a
  gravitational collapse similar to the $\alpha_d=0$ case studied earlier
  (Figures  \ref{fig:fixed_a10}).}
\label{fig:fixed_a30_alpha1m3}
\end{figure}

Our final experiment with fixed size grains is to repeat the calculation of \S
\ref{sec:fixed_a_mid_turb} but with a larger size of grains, $a = 30$
cm. Figure \ref{fig:fixed_a30_alpha1m3} demonstrates a very rapid
sedimentation of grains in this case, with gravitational collapse of the
``grain cluster'' (\S \ref{sec:grain_cluster}) occurring by about 500 years. As
expected, larger grains are affected significantly less by the turbulent
mixing.

The rapid grain sedimentation in this calculation yields a mean grain
sedimentation velocity of about $15$ m s$^{-1}$ in this model. Such a rapid
sedimentation would undoubtedly lead to shattering of some of the grains, so
that this calculation is somewhat unrealistic, but it goes to show that if
grains can grow to decimetre sizes, then they can probably overcome the
effects of turbulent mixing.

\subsection{Full models}\label{sec:a1m2}

\subsubsection{Simulations Table}\label{sec:table1}

We shall now explore our ``full'' model that includes the grain growth
prescriptions (\S \ref{sec:num_ggrowth}). Selected runs that we performed and
discuss below are listed in Table \ref{table1}. Some of the parameters of the
models have very little bearing on the results, except near the boundaries
separating qualitatively different regimes. These parameters are not listed in
the Table. In particular, the results are insensitive to the initial size of
the grain, $a_0$, and the Brownian motion velocity, $v_{\rm br}$, as long as
the latter is large enough (most of the runs were performed with $v_{\rm br} =
5$ cm s$^{-1}$). This is because, starting from $\mu$m-size grains, one finds
that grains quickly reach sizes of hundreds of $\mu$m anyway \citep[see also
  more detailed calculations, showing a similar initial rapid grain growth,
  by][]{DD05}. Thus $a_0$ is set to $100 \mu$m for most tests below.

Given the large differences in the cooling times (\S
\ref{sec:1st_analytical}), the form of the opacity law predictably turns out
to be quite important. The names for the runs performed with the power-law
index $\alpha=1$ starts with a letter ``S'' for ``small opacity'', whereas
names for $\alpha=2$ runs start with ``L'' for ``large opacity''. These names
are not to be taken literally as there is also the opacity coefficient,
$\kappa_0$, in front of the opacity law we use here (equation
\ref{kappa0}). $\kappa_0$ is also varied in some models.

The last three columns in table \ref{table1} list several output quantities of
the simulations. $t_{\rm grain}$ , $T_c$ and $\rho_c$ are the time at which
the self-gravitating phase of the grain sphere evolution is reached (if ever),
and the central temperature and the density of gas, respectively, at that
time. Grains were vaporised before they could make a massive self-gravitating
sphere in the centre in the runs in Table 1 for which no values of these three
parameters are given.

\subsubsection{Turbulent mixing versus grain growth}\label{sec:turb_vs_growth}

We first discuss runs L1, L2 and L3, the three models that are identical to
each other except for the value of the turbulent mixing coefficient, which was
set to $\alpha_d = 10^{-4}, 10^{-3}$ and $10^{-2}$, respectively. There is
very little difference in the evolution of the models. This may appear
surprising given the all-important role the turbulence played in the
simulations with the fixed grain size. The ``paradox'' is resolved by
realising that the runs always start with the turbulence-dominated regime
(small grains) and end up in the negligible turbulence one (large grains), and
only the time and the boundary between these two vary in L1-L3.

\begin{figure}
\centerline{\psfig{file=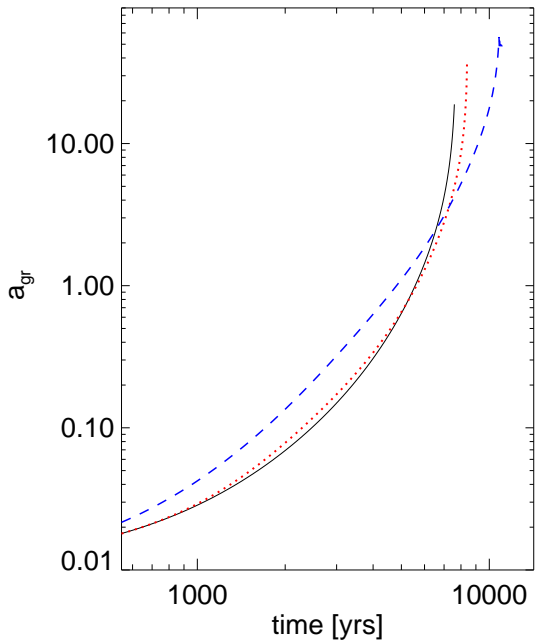,width=0.45\textwidth,angle=0}}
\caption{Grain size versus time for models described in \S \ref{sec:a1m2}. The
solid black, the red dotted and the blue dashed curves are for $\alpha_d = 10^{-4},
10^{-3}$ and $10^{-2}$, respectively.}
\label{fig:a_vs_t}
\end{figure}

Figure \ref{fig:a_vs_t} shows the grain size, $a$, versus time for these three
models. The curves are terminated at the point when the central radial bin is
dominated by the grains. Initially the curves in the figure are nearly
identical to each other. This can be understood by realising that at a small
$a$, sedimentation speeds are low, so that turbulent mixing easily balances
sedimentation. A quasi-steady state results in which grain sedimentation is
neutralised by turbulent mixing. 

Figure \ref{fig:a_vs_t} shows that grain growth is slightly faster for higher
levels of turbulence. We explain it by the fact that the grain distribution is
slightly more extended in this case, hence the sedimentation velocity,
proportional to $R$ (see equation \ref{vsed}), is larger, allowing the grains
to sweep up smaller ones at a faster rate and hence grow faster.

The process of grain growth accelerates with time. At early times this is due
to the fact that $u_{\rm sed} \propto a$, thus $da/dt \propto a$ (see equation
\ref{dadt2}). At later times, grains become over-abundant in the inner
regions, i.e., $\rho_g/\rho > f_g$ there. This speeds up the rate at which the
larger grains can sweep up the smaller grains even further.  This non-linear
stage of grain growth is naturally reached faster by the lower turbulence
models. A corollary of this is that the higher turbulence runs reach the grain
self-gravitating phase with higher temperature and density.

Nevertheless, it is notable how qualitatively similar the three simulations
actually are. One difference is the time when the central region becomes
grain-dominated. It is $t=7.6$, 8.4 and 11 thousand years for $\alpha_d =
10^{-4}, 10^{-3}$ and $10^{-2}$, respectively. The later the grain-dominated
phase reached, the hotter is the central region of the first core, and the
denser it is (see Table 1).

The main conclusion we can draw from these experiments is that the turbulent
diffusion is only likely to delay rather than stop grain sedimentation {\em
  if} grain growth can proceed via the hit-and-stick process faster than the
gas heats up to vaporise the grains.

\subsubsection{Delayed sedimentation due to grain
  fragmentation}\label{sec:delayed}

To explore sensitivity of our grain growth model to grain fragmentation due to
high speed collisions, we ran several models where the maximum velocity
parameter $v_{\rm max}$ was reduced from the nominal value of $10$ m s$^{-1}$
to 1 m s$^{-1}$ (see Table \ref{table1}). In particular, simulation L2a is
identical to L2 except for the smaller value of $v_{\rm max}$. Qualitatively
the two runs are similar, but the grain-dominated cluster phase is reached at
time almost twice as long in L2a than in L2. This is caused by the subdued
grain growth rate in the run L2a compared with L2. 

Concluding, as envisaged earlier in \S \ref{sec:destr_coll}, grain
sedimentation may still proceed reasonably quickly at realistically small
($\sim $m s$^{-1}$) values of $v_{\rm max}$.

\subsection{Vaporisation of grains in rapidly cooling first
  cores}\label{sec:too_hot} 

In the context of our model of isolated first cores, the only robust way to
stop grain sedimentation is to prevent their growth. This occurs when the
first cores become too hot, so that grains vaporise. We find this behaviour if
(a) the cores are too hot (too massive) to begin with; (b) the opacity is low
so that the cores cool very quickly (thus leading to hotter internal
temperatures); (c) the grain content of the core is low, yielding too long
grain growth times.

\begin{figure}
\centerline{\psfig{file=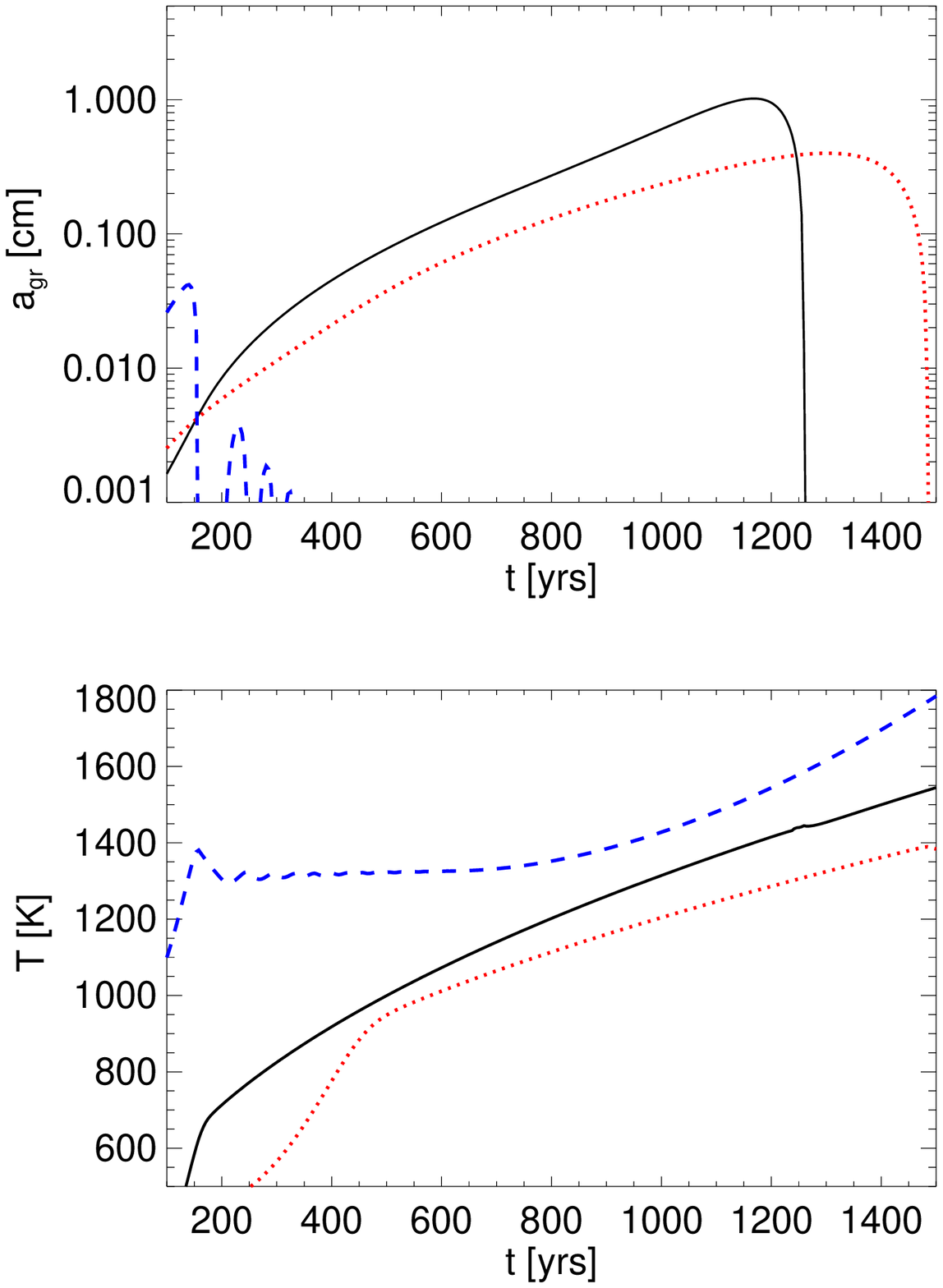,width=0.45\textwidth,angle=0}}
\caption{Grain size versus time (upper panel) and the central gas temperature
  (lower panel) for three cases that led to grain vaporisation rather than
  growth: S3 (solid black), S6 (dashed blue) and S7 (red dotted). In all three
  cases the first core cools radiatively, contracts, and becomes too hot to
  allow grain growth.}
\label{fig:no_sed}
\end{figure}

Figure \ref{fig:no_sed} shows the grain size and temperature as a function of
time for three runs that yielded no grain sedimentation. The solid black curve
is the simulation S3 (cf. Table 1) -- a low opacity, $\kappa_0 = 0.001$, and
low grain mass fraction case, $f_g = 1.2 \times 10^{-3}$, with the first core
mass of $\mfc = 5 M_J$. The opacity law exponent is $\alpha=1$. Both
conditions (b) and (c) are the case for this particular run. The grains manage
to reach the size of about 1 cm before they are vaporised. The central gas
temperature is almost $1400$ K at that time. Grain vaporisation actually
starts earlier, at $T_c \sim 1300$ K or so, but does not immediately prevail
over the process of grain growth.

The other two runs shown in figure \ref{fig:no_sed} have much more massive
cores, $\mfc = 25 M_J$ (S7 in Table 1; dotted red curve) and $\mfc = 20 M_J$
(S6; dashed blue). The opacity coefficient $k_0 = 0.05$ and $0.01$, for the
runs, respectively, and the grain mass fraction is $f_g = 0.01$ for both. The
cores cool quickly nevertheless as the opacity power law index $\alpha=1$,
hence the short cooling times (equation \ref{tcool1}). The most massive first
core considered here (blue curve) is almost too hot to allow grain growth to
begin with (case ``a'' above). As the cooling time is short, the core
undergoes a quick and nearly dynamical evolution in the beginning, driving
hydrodynamical oscillations that allow some early grain growth. However the
latter is quickly turned into grain evaporation as the core settles into its
cooling evolution.

\subsection{Importance of grain fraction (metalicity)}

Amongst the parameters of the simulations that we varied is the first core's
initial grain mass fraction, $f_g$ (Table 1). A higher grain content obviously
favors grain growth. For example, the runs L4 and L5 are identical except for
the grain mass fraction, $f_g = 0.005$ and $0.02$, respectively. Not
surprisingly, the grain-dominated phase is reached far faster in L5 than in
L2. Therefore, we can expect that metalicity of the first core is one of a
key parameters determining whether grain sedimentation takes place or not.

\begin{table*}
\caption{Parameters of the simulations. The mass of the first core, $\mfc$, is
  in Jupiter's masses; $k_0$ is the opacity coefficient in the opacity law
  given by equation \ref{kappa0}; $\alpha$ is the power law index in that law;
  $f_g$ is the grain mass fraction of the core; $\alpha_d$ is the turbulent
  mixing coefficient; $v_{\rm max}$ [m s$^{-1}$], the maximum velocity before
  grain-grain collisions stifle grain growth (equation \ref{dadt2}); $t_{\rm
  grain}$ is the time at which the innermost gaseous cell becomes dominated by
grains, by mass, in units of $10^3$ yrs; and $T_{\rm c}$ and $\rho_{\rm c}$
are the gas temperature and density there at the time, in cgs units.  }
\begin{center}
\begin{tabular}{|c|c|c|c|c|c|c|c|c|c|}\hline
ID & $\mfc$ & $k_0$ & $\alpha$ & $f_g$ & $\alpha_d$ & $v_{\rm max}$& $t_{\rm grain}$ (10$^3$
yrs) & $T_{\rm c}$ & $\rho_{\rm c}$
\\
                                  \hline
S1 & 10  & 0.01 & 1 & 0.005 & $5\times 10^{-2}$ & 10 & $2.7$ & 1400 & $1.4 \times 10^{-8}$ \\ \hline
S2 & 10  & 0.001 & 1 & 0.005 & $5\times 10^{-2}$ & -- & -- & --\\ \hline
S3 & 5  & 0.001 & 1 & 0.0012 & $10^{-3}$ & 1 & -- & -- & --\\ \hline
S4 & 5  & 0.003 & 1 & 0.0012 & $10^{-3}$ &1 & 3.3 & 1270 & $1.5
\times10^{-8}$\\ \hline
S6 & 20  & 0.01 & 1 & 0.02 & $10^{-3}$ &1 & -- & -- & --\\ \hline
S7 & 25  & 0.05 & 1 & 0.02 & $10^{-3}$ &1 & -- & -- & --\\ \hline
L1 & 10  & 0.01 & 2 & 0.005 & $10^{-4}$ & 10 &7.8  & 695 & $7.5 \times10^{-10}$  \\ \hline 
L2 & 10  & 0.01 & 2 & 0.005 & $10^{-3}$ & 10 &8.4  & 710 & $8 \times 10^{-10}$\\ \hline
L3 & 10  & 0.01 & 2 & 0.005 & $10^{-2}$ & 10 &11.0 & 750 & $10^{-9}$ \\ \hline 
L2a & 10  & 0.01 & 2 & 0.005 & $10^{-3}$ & 1 &13.4  & 790 & $1.2 \times
10^{-9}$\\ \hline
L4 & 10  & 0.001 & 2 & 0.005 & $5\times 10^{-2}$ & 10 &1.9 & 1190 & $4 \times 10^{-9}$ \\ \hline 
L5 & 10  & 0.001 & 2 & 0.02 & $5\times 10^{-2}$ &10  &0.69 & 1130 & $3 \times
10^{-9}$ \\ \hline 
L6 & 10  & 0.1 & 2 & 0.005 & $10^{-3}$ &10 &16.0 & 360 & $1.3 \times 10^{-10}$ \\ \hline 
L7 & 40  & 0.1 & 2 & 0.005 & $10^{-3}$ &10 &15.3 & 1100 & $2.1 \times 10^{-10}$\\ \hline
L8 & 55  & 0.1 & 2 & 0.005 & $10^{-3}$ &10 & -- & -- & -- \\ \hline
\end{tabular}
\end{center}
\label{table1}
\end{table*}

\section{Discussion}\label{sec:discussion}

\subsection{Summary of the results}

In this paper we explored the possibility that grains may sediment inside
isolated or slowly accreting first cores, which are the gaseous condensations
with mass between about 5 and 50 Jupiter masses. While many parameters (such
as the turbulent viscosity parameter) determine the end result, the main
requirement for sedimentation to take place is a rapid enough grain
growth. Based on numerical experiments, sedimentation occurs as long as grains
can grow to a few cm to a few tens of cm before vaporisation temperature of
$T\approx 1400$ K is reached. This is expressed mathematically as
\begin{equation}
t_{\rm gr} < t_{\rm vap}\;,
\label{time_cond}
\end{equation}
where $t_{\rm gr}$ is the grain growth time scale (equation \ref{tgr_exact}),
and $t_{\rm vap}$ is the vaporisation time of the grains (equation
\ref{tvap_in}). Turbulent mixing and grain shattering by collisions may slow down
grain growth compared with the analytical model (in which case $t_{\rm gr}$
must be obtained numerically).

While the outcome depends on several parameters -- mass of the first core,
$\mfc$, ambient gas temperature, $T_{\rm init}$, opacity law in the first
core, the initial grain mass fraction, $f_g$, turbulent mixing, etc., we shall
here give only one example of the implications of the equation
\ref{time_cond}. In particular, we shall fix $f_g = 0.005$,
$\alpha_d=10^{-3}$, $v_{\rm max} = 10$ m s$^{-1}$, and $T_{\rm init} = 10$ K. 

With these parameters fixed, we can ask the following question: At a given
opacity coefficient $\kappa_0$, what is the range of the first core masses
that could support the dust sedimentation?

\begin{figure}
\centerline{\psfig{file=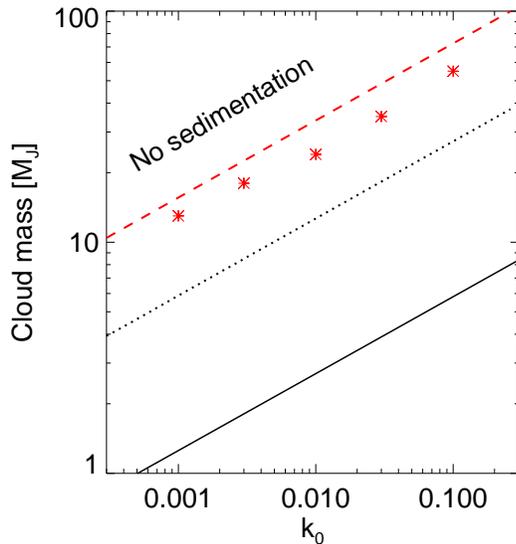,width=0.45\textwidth,angle=0}}
\caption{Minimum first core mass (solid curve) and maximum first core mass for
  grain sedimentation versus the opacity coefficient $\kappa_0$ for $\alpha=1$
  (black dotted line) and $\alpha=2$ (red dashed line). No grain sedimentation
  is expected above these lines as grains get vaporised faster than they can
  sediment.  See text for further detail.}
\label{fig:mcloud_max}
\end{figure}

Figure \ref{fig:mcloud_max} answers this question for the two opacity
power-law indexes that we studied here, $\alpha=1$ (black dotted curve) and
$\alpha=2$ (red dashed one). The lower (solid) line gives the minimum mass of the
first core which we obtained by requiring that the mean first core density
(equation \ref{rho_mean}) is at least equal that at which the collapsing gas
switches to the adiabatic behaviour (equation \ref{mi99}). This curve is
independent of $\alpha$ for $T_1 = 1$. The curves dividing the parameter space
are obtained by solving the equation \ref{time_cond} with our analytical model
for grain growth and the first core contraction. The asterisks are the maximum
first core mass for which grain sedimentation occurs found numerically for the
$\alpha=2$ case. They are encouragingly close to the analytical curve. The
fact that they are slightly below is probably explained by the extra delay in
the grain growth that the turbulent mixing and the maximum velocity $v_{\rm
  max}$ impose in numerical simulations compared with the analytical models that
do not account for these processes.

This limited survey of parameter space demonstrates that there range in $\mfc$
in which grain sedimentation is possible is about a factor of 10 wide for
$\alpha=2$, and shifts to higher masses for higher opacities. This is to be
expected, as higher opacity implies longer cooling times.

The $\alpha=1$ case (black dotted line) shows a narrower window of opportunity
for dust sedimentation, which is again a cooling effect. As these cores cool
quicker, there is less time for dust growth.

\subsection{Comparison to previous work}\label{sec:comp} 

Our analytical model and numerical simulations confirm the suggestion made by
\cite{Boss98} that grains can grow and sediment inside giant embryo, although
started from a much less dense initial configuration. While a fuller
comparison is to be made in paper II where we continue the calculations to the
point of the solid core formation, we can already see that this lower density
and temperature initial configuration is key to address the criticism levelled
on the \cite{Boss98} model by \cite{WuchterlEtal00}, later confirmed by
\cite{HelledEtal08,HS08}. These authors found that their giant planet models
were convective, and that convective grain mixing has significantly slowed
down the core growth.

Since the initial column depth of the embryos is much lower in our models, the
radiative cooling remains strongly dominant over convective cooling for $\sim
10^4 - 10^5$ years, the time it takes (see figure 2) for the embryos to
contract to densities and temperatures considered by the authors mentioned
above. These findings echo the results by \cite{Bodenheimer74} who studied
early contraction of a 1 Jupiter mass gas cloud and concluded that the planet
remained radiative during the first $\sim 10^5$ years. While his model does
not include dust sedimentation, it accounts for H$_2$ molecules
dissasociation, hydrogen ionisation, and much more (that we do not include
here). His calculations also used similarly cool initial conditions ($43$ K
was the initial temperature of his model planet). 

Therefore it appears that most of the differences between us and
\cite{HelledEtal08,HS08} can be traced to the significantly different initial
conditions.  In terms of applications to planet formation field, the
differences in our results and that of \cite{HelledEtal08,HS08} are
striking. These authors found that a 5 Jupiter mass embryo cannot yield any
dust sedimentation as it is too hot to begin with, e.g., the initial
temperature above the grain vaporisation temperature. In contrast, we find
that a 5 Jupiter mass embryo provides an excellent environment for dust
growth: the gas temperature is initially less than 100 K and stays below 1400
K for as long as $10^4$ to more than $10^5$ years, depending on the opacity
(cf. figure 1).

In paper II we do find that convection becomes very important in the inner
region of the embryo {\em once a massive solid core forms}, as this releases a
significant amount of heat. It is also interesting to note that while our
initial condition is significantly different, the inferred grain growth and
sedimentation time scales are from $10^3$ to $10^4$ years, in a broad
agreement with the estimate by \cite{Boss98}.

On the other hand, in a qualitative agreement with \cite{HelledEtal08,HS08},
the conditions for dust growth found to deteriorate with an increasing mass of
the gas clumps. Depending on opacity, grains cannot sediment for giant embryo
masses higher than $\sim 10$ to a few tens Jupiter masses.

\subsection{Astrophysical implications}\label{sec:implications}

Our calculations here and in paper II lend support to the ideas presented by
\cite{Boss98}, except that grain sedimentation process must start earlier in
the life of the protoplanet, while it can still be considered the first
core. Lower initial temperatures and rather long cooling times of the giant
embryo in the latter stage are key for a successful grain sedimentation
outcome.

\cite{BossEtal02} argued that giant embryos may yield not only giant planets
but also giant icy planets with cores if the metal poor envelopes of the
embryos are removed. Irradiation by a nearby star was suggested to accomplish
this. While this is certainly possible, it may be not very probable as OB
stars are rare.

The present paper is the base for paper III, where we argue that giant embryos
may migrate inward due to the gravitational torques from the disc.  Estimates
show that embryos may migrate to the distance of several AU to the parent star
in several $\times 10^4$ years, typically. As cooling of embryos slows down as
they age, the embryos must be eventually disrupted by the tidal forces or by
the heating due to irradiation from the parent star. This opens up an exciting
possibility that {\em all} planets may be formed by such a modified version of
gravitational instability model.

\section{Conclusions}

In this paper we considered the grain growth and sedimentation process inside
the giant planet embryos, which we argued must be nearly identical in
properties to the first cores. These are the first gaseous hydrostatic
condensations from which stars may form. We found that grains can indeed grow
and sediment to the centre of the gas cloud provided that the gas remains cool
enough (temperature below $\sim 1400$ K) for the time it takes the grains to
reach a few cm size. The efficiency of the dust sedimentation process and the
final mass of the solid core are strong functions of opacity and other
parameters of the problem.

We suggested that astrophysical applications of these results may be in the
field of planet formation, where giant planet embryos may serve as birth
places for all types of planets if these embryos migrate inward and get
tidally or irradiatively disrupted in the inner few AU from the parent star.

\section{Acknowledgments}

The author is indebted to Richard Alexander and Seung-Hoon Cha for useful
discussions of the problem, and to Roman Rafikov and Phil Armitage for
comments on the draft.  Comments by the anonymous referee were instrumental in
improving the clarity and astrophysical relevance of the paper. Theoretical
astrophysics research at the University of Leicester is supported by a STFC
Rolling grant.


\label{lastpage}

\end{document}